\documentclass [12pt]{article}

\usepackage{float}
\usepackage{natbib}

\usepackage{graphicx} % Required for inserting images
\usepackage{amsmath}
\usepackage{amssymb}
\usepackage{placeins}
\usepackage{booktabs}
\usepackage[utf8]{inputenc}
\usepackage{caption}
\usepackage{indentfirst}
\usepackage{float}
\usepackage{siunitx}

\title{Magnetic Centrifuge Effects in Ultrafast Laser Ablation Plasmas}%

\author{
Peter P. Pronko$^{1}$ and Paul A. VanRompay$^{2}$
}

\date{}

\begin{document}

\maketitle

\begin{center}
$^{1}$Emeritus Research Scientist, Department of Electrical Engineering and Computer Science,\\
University of Michigan, Ann Arbor, Michigan, USA\\
\texttt{pronko@umich.edu}

\vspace{6pt}

$^{2}$NOAA National Oceanic and Atmospheric Administration,\\
Satellite Communications Division\\
\texttt{pvanrompay@gmail.com}
\end{center}

\pagestyle{plain}
\pagenumbering{arabic}

\pagestyle{plain}
\pagenumbering{arabic}
\maketitle
\begin{abstract}
 A self-consistent model is developed to explain the anomalously large enrichment of nickel isotopes observed in ablation plumes from ultrafast laser irradiation of solid surfaces. The model is based on the spontaneous creation of a magnetic centrifuge in the ablation plume and the associated cyclotron rotation of plasma ions with effective rotation rates on the order of $10^9$ radians per second. Mass separation occurs around the radial coordinate of cylindrical symmetry with longitudinal axis normal to the ablating surface. A Gaussian shaped radial magnetic field $B_{eff}$ is extracted for Ni isotopes which is shown to be a combination of an axial $B_z$ component and a second contribution $B_{ibw}$ that represents the equivalent of an “effective magnetic field” contributing to the isotopic separation due to broad spectrum Ion Bernstein Waves providing electrostatic acceleration to the cyclotron orbits. These IBWs are also responsible for a profound resonance of enrichment observed for certain specific charge states.   In addition to cyclotron rotation of ions, a rigid rotor model is also presented that is associated with the hydrodynamic rotation of the entire plasma and is shown to be of little consequence for the isotope enrichment. Cyclotron rotation and IBWs dominate the process.
\end{abstract}

\section{Introduction}
\indent
    In a paper published in 1999 we reported unusually strong isotope enrichment in the normal ejection direction from ultrafast femtosecond laser ablation pulses using laser power densities in the $10^{13}-10^{15}$ W/cm2 \cite{pronko1999isotope}. It was postulated, at that time, based on the angular distribution of those enrichment and concentration distributions of deposited material, that an intense Megagauss magnetic field was responsible for the observed effects. Since then, a number of papers have appeared in the literature reporting on experimental observations of such fields \cite{kahaly2009polarimetric, pisarczyk2015space, pisarczyk2017kinetic,sandhu2002laser} . In view of these developments and the ongoing interest in such magnetic fields, we are revisiting our earlier work and providing important additional data that was obtained during that period, and that has hitherto not been published but which bear directly on the subject\cite{van2003mass}. 
    At the time of that earlier 1999 paper, we were criticized for proposing a magnetic field effect for our observed results, with alternative explanations proposing hydrodynamic separation as being responsible for the observed phenomena \cite{gupta2001comment}. That criticism was leveled at our work out of an incomplete understanding of the characteristics of ultrafast ablation plumes and the various time-dependent processes that occur during its temporal development. In addition, thermodynamic arguments that are relevant to nanosecond laser ablation were being referenced and used in those criticisms. The difference between femtosecond and nanosecond laser ablation pulses is now known to be associated with the fact that femtosecond ablation plumes are characterized by, among other things, a Coulomb explosion emanating out of a solid density plasma \cite{vanrompay1998pulse} and not by the thermodynamic plasma formation associated with nanosecond ablation plasmas. In further support of a centrifuge process it is noted that hydrodynamic separation would show a $1/\sqrt{m}$ mass dependence whereas we observe a mass difference  $\Delta m$ dependence as predicted by a centrifuge process \cite{vanrompay2000isotope}. In the present paper, we will present data showing that the elemental and isotopic spatial separations can be consistently described by a magnetic centrifuge process occurring in the self-generated magnetic fields of these ultrafast ablation plasmas. Further we shall show that the magnetic centrifuge can be described  by specific Larmor orbits of ions within the plasma with assistance from resonant and broad spectrum ion Bernstein waves (IBW). Details concerning resonant separation enhancement of specific charge states by these internal wave excitations is presented and discussed. Rigid rotor plasma rotation is demonstrated to be of no significance in the isotope enrichment process.

    \subsection{Ablation Dynamics}
The generation of a plasma from a solid surface by a laser pulse involves the absorption of photon energy by the electrons during the time of the laser pulse and the transfer of that energy to the atoms and ions in the material structure. This transfer of energy to the atoms ultimately results in an ablation plume emanating from the surface. The process of that energy transfer is specific to the laser pulse characteristics and the internal dynamics of the plasma formation. It is important to recognize here that the ablation plume, which is emitted normal to the target surface, will have both spacial and temporal characteristics. We examine the spacial distribution of ions in this report but need to recognize that the plume develops not only with ions, but also with neutrals and nano-particles. This occurs along a specific time line where the pulse absorption is characterized initially by a solid density plasma with fast and slow ions followed by neutrals, and subsequently followed by a burst of nano-particles \cite{albert2003time}. These events are observed to have individual time horizons and can, in principle, be separated and studied using mechanical or electrostatic separation mechanisms. Harvesting isotopic ions suggests itself in the form of electrostatic separation.

In regard to the initial plasma state, it is considered axiomatic that a plasma will self-organize after being formed as it relaxes towards a quasi- equilibrium transitional state before eventually dispersing \cite{boyd2003physics,bittencourt2013fundamentals}. The specific ways this happens will depend on factors such as the electro-magnetic field details that are present and the characteristics of the plasma itself.
In the case of high intensity ultrafast laser pulses, as used in this work, the pulse time scales are femtoseconds and intensities on the order of $10^{14}$ to $10^{15}$ watts/cm$^2$. Due to the short time scales and the rapid absorption of energy by the lighter electrons, a coulomb explosion occurs at the surface whereby hot electrons which are ejected outward create a charge imbalance leaving a positive surface layer of ions that are accelerated outward giving them energies in the keV range (see Appendix 8.1) \cite{vanrompay1998pulse}. This explosion and transfer of energy happens over a time scale on the order of a few picoseconds or less.  During this time interval, electrons also circulate above the target surface and along the low density plasma sheath, acting together with internal thermoelectric currents to form toroidal and longitudinal magnetic fields within the plasma.

The expansion of the initial solid density plasma can be considered to evolve as a one-dimensional (1-D) process, moving as an expanding plug along the normal to the surface of the ablating material \cite{puell1970heating}.  The conversion to a three-dimensional (3-D) expansion occurs after a distance that is on the order of or somewhat greater than the laser beam spot size \cite{stevefelt1991modelling}.  Within the one-dimensional region of the expanding plasma, both toroidal and longitudinal magnetic fields of very high intensity have been observed and measured \cite{kahaly2009polarimetric, pisarczyk2015space,sandhu2002laser}.  

The toroidal, or azimuthal, field is considered to form first from rapid outflow of hot electrons under the influence of such mechanisms as the Biermann battery effect. Subsequently a longitudinal field along the one-dimensional z axis forms due to a dynamo effect from a radially expanding, slightly rotating plume that will naturally convert some of a strong toroidal field $B_{\theta}$ into an axial (longitudinal) field $B_{z}$. This occurs primarily through the MHD induction term $\nabla \times (\mathbf{v} \times \mathbf{B})$ together with turbulence such as rotor rotation, tangential surface loops, and non-symmetric three dimensional charge expansion to establish, within picoseconds, a well defined longitudinal field \cite{briand1985axial}.  In addition, the toroidal field will continue to act producing a magnetic pinch on the ionic component of the plasma \cite{yutong2000observation,jackson2021classical}.  This pinching effect will work to hold the plasma together within the one-dimensional zone. The longitudinal magnetic field induces a rotational motion to the expanding plasma and to the energetic particles within, occurring through transverse velocity components and drift rotation, similar to the way it occurs in a conventional plasma centrifuge device \cite{krishnan1981plasma}.  The combined effect of these two phenomena, pinching and rotation, can be expected to induce a radial separation of ionic masses within the 1-D cylindrical region of the expansion zone.  To the extent that the distribution of ions in the cylindrical zone reaches some quasi equilibrium state, it is expected that these rotational processes
will achieve a centrifuge type distribution \cite{krishnan1981plasma}. The plasma behavior in magnetic centrifuge devices has been studied in considerable detail.  For the case of an electro-magnetically-driven plasma centrifuge, the rotation rate can be one of two distinct angular frequencies, one at the cyclotron frequency, and one at a specific lower rate, proportional to the $\vec{E} \times \vec{B}$ drift rotation.  Derivation and discussion of these two rotational mechanisms are provided in references \cite{kim1987equilibria,gueroult2019b}.  In conventional rigid-rotor plasma devices, only the lower of the two characteristic rotation rates is usually observed.  This most probably is a consequence of the relatively low magnetic field strengths($ \approx 5$kG) used in those devices.  In contrast, we show below that our data demonstrates ion rotations at cyclotron frequencies, in these ultrafast ablation plasmas, where much higher field strengths are present. 

When the ions move into the three-dimensional expansion region, they will transform out of cylindrical symmetry and into a divergent conical symmetry characteristic of pulsed-laser plasma expansion.  Figure 1 shows an idealized behavior for how these ions could be considered to undergo this geometric transformation process.  The figure will also be used to discuss model calculations presented herein.  Those ions on the central axis of the cylinder are considered to remain close to the central axis of the expansion cone, whereas those on the extreme edge of the cylindrical radius are expected to expand out to the largest angles of the cone.

\subsection{Laser Parameters and Detector-Target Geometry}
The laser parameters are 10~Hz repetition rate at 780~nm wavelength with a pulse width of 185~fs (FWHM) and a contrast ratio of $10^6$. The beam is $P$-polarized and focused to a 50~$\mu$m FWHM spot. The focused laser intensity on target is $5.2 \times 10^{15}~\mathrm{W/cm^2}$ using a pulse energy of $13~\mathrm{mJ}$. The laser beam is incident on the target at 45 degrees and the ablation plume leaves normal to the surface. As will be discussed later, p polarization and oblique incidence is critically important for these studies. The detector is fixed in position and the sample is scanned from zero to 60 degrees with the plume always leaving normal to the surface. \cite{vanrompay2000isotope,van2003mass,usPatent6787723B2}. It is important to include here some comments about the experimental set up of the ion spherical sector analyzer that was used in these experiments. It can be shown that it operates effectively as an ion-plasma-microscope seeing only a very small fractional area of the plasma at any given setting. The detector is 1.1 meters away from the ablation surface. It has a 2 mm entrance aperture. In addition to this there are two 4 mm apertures placed 1.5 inches apart and centered at 5 inches upstream of the detector entrance. All of this strict collimation results in the detector seeing only a very small selected region of the expanding cylinder end of the emitted plasma ions. In addition, the conical expansion of the initial cylinder produces a magnification effect further increasing the microscope character of the detector. It can be shown (see Appendix 11.2) that for the this arrangement, the visible area of the plasma at the detector is only about 0.22 microns in diameter out of the full 125 micron radial zone of the ablation end cap. In this configuration it provides extraordinary spatial resolution on the ion mass concentrations, charge states, and energy distributions. For details concerning this experimental set up see \cite{vanrompay2000isotope,van2003mass, usPatent6787723B2} and Appendix Sec. 11.2.

\section{Equations and Data}
The generalized centrifuge equation, valid for cylindrically symmetric isotopic separation in a mechanical gas centrifuge or rotating-plasma device, is given by:
\begin{equation}
\left(\frac{n_2}{n_1}\right)_r= \left(\frac{n_2}{n_1}\right)_0\exp\left[\frac{\Delta m (\omega \ r)^2}{2 k_B T}\right ] 
\end{equation}
which expresses the mass fraction of a Boltzmann energy distribution within the cylinder. The terms in this Gaussian function are:
$n_2/n_1$ = concentration ratio of mass $m_2$ and $m_1$ at radius r and at 0 (on axis), $k_B$ is the Boltzmann constant, T the plasma ion temperature, $\omega$ the centrifuge angular rotation rate and r 
the radial extent of the rotating mass.

\
Figure 1 shows the essential features of the geometry
being used for coordinate transition from an initially expanding cylindrical plug of plasma at the point of laser impact to its subsequent expansion into a conical ablation plume.The laser spot diameter is noted as $d_1$ whereas $d_p$ is the plasma plume diameter. For clarity, the 1-D zone has been expanded; the length is approximately $d_p$/2. The coordinate transformation in Fig.1 shows the relationship between the cylindrical geometry of the plasma centrifuge (with mass separation by radius r) and the angular distribution associated with the ablation plasma (with separation mapped into angle $\theta$). Transforming r to $\theta$, an elementary conformal mapping function is used
such that $r/r_{max}$ = $\theta/\theta_{max}$.

\begin{figure}[t]
\vspace*{-3.0cm}
\hspace*{-2.5cm}
  \includegraphics[width= 1.2\textwidth, trim=0 0 0 0, clip]{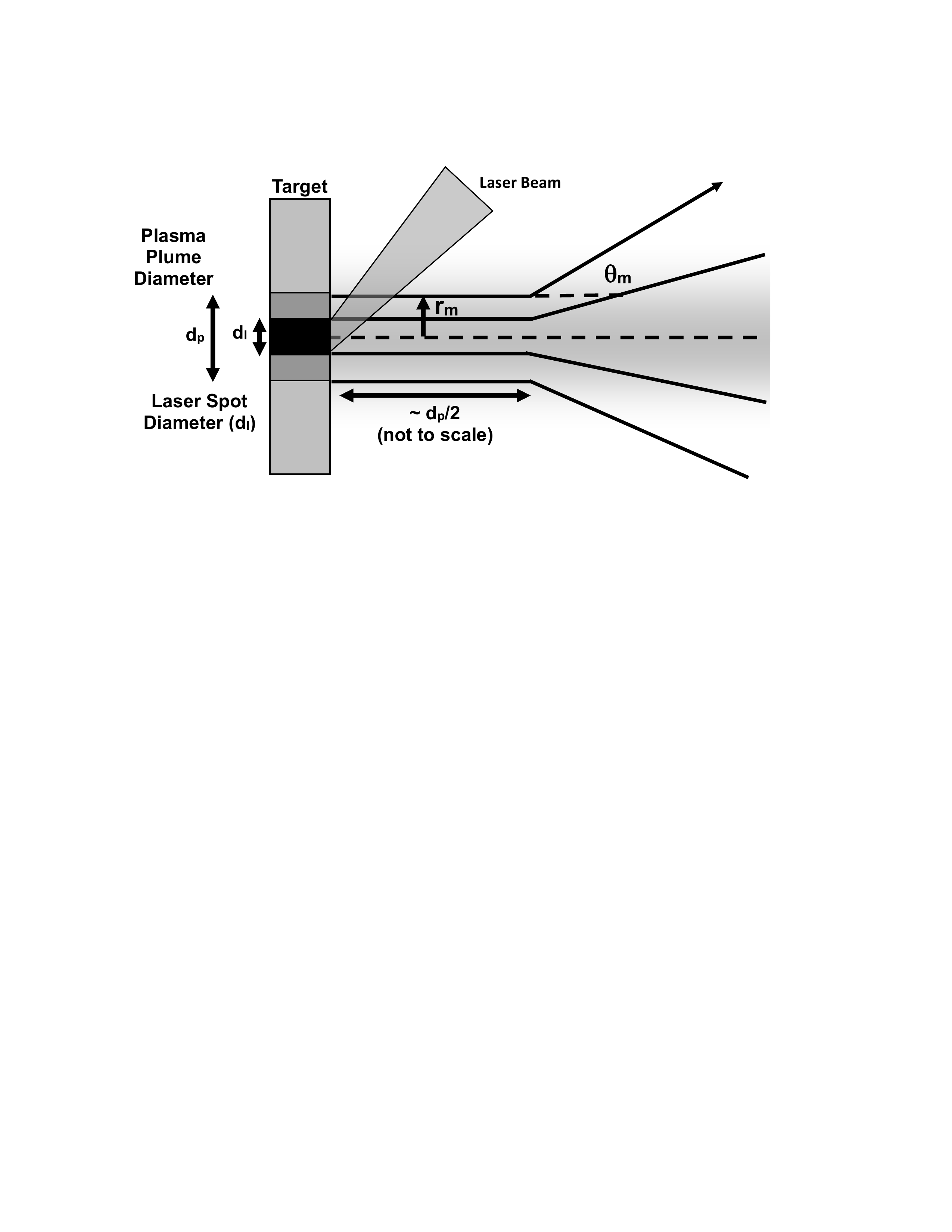}
   \vspace{-350pt}
    \caption{Ablation Model for Enrichment Data as a Function of Observation Angle $\theta$. In the present work $r_m$ is taken as 125 microns which is 1.5 times the 50 $\mu\text{m}$ radius of the $(1/e^2)$ laser intensity profile} 
    \label{fig:fig2}
\end{figure}

Defining a separation factor S(r) = $(n_2/n_1)_r/(n_2/n_1)_0$
and performing the \ensuremath{\theta} coordinate transformation, the basic centrifuge equation is converted to angular coordinates as shown in Equation 2.
\begin{equation}
S(\theta) = \exp \left[ \frac{\Delta m\omega^2 r_{\max}^2}{2 k_B T \theta_{\max}^2} \theta^2  \right]
\end{equation}
\vspace{10pt}

\subsection{Data}
\indent
Two different ablation experiments are presented below.
An experiment to test the ablative separation of chemical elements has been performed on a Ni/Cu alloy of 45\%Ni and 55\% Cu. Figure 2 shows the distribution of the Ni/Cu element separation ratios as a function of charge state and angle for the specific angles of 0, 15. 30, and 60 degrees. Charge states +1 through +10 are shown as gray bars, with the highest enrichment labeled. 
The total enrichment ratio for all ions is plotted as circles. The enrichment ratio is normalized to the element composition ratio. The gray bars represent the contribution of the various charge states from +1 to +10. These data are presented as clusters at the specific angles of 0, 15 ,30, and 60. It is noted that each group has a dominantly enriched charge state along with sub-dominant peaks. These will be discussed further below. Figure 3 shows a similar plot of the $Ni^{58}/Ni^{60}$ isotope ratios, obtained from a pure nickel target, also showing especially strong dominance of particular charge state enrichments.

\begin{figure}[H]
\vspace*{-3.0cm}
\hspace*{-5.0cm}
  \includegraphics[width= 1.4\textwidth, trim=10 0 0 0, clip]{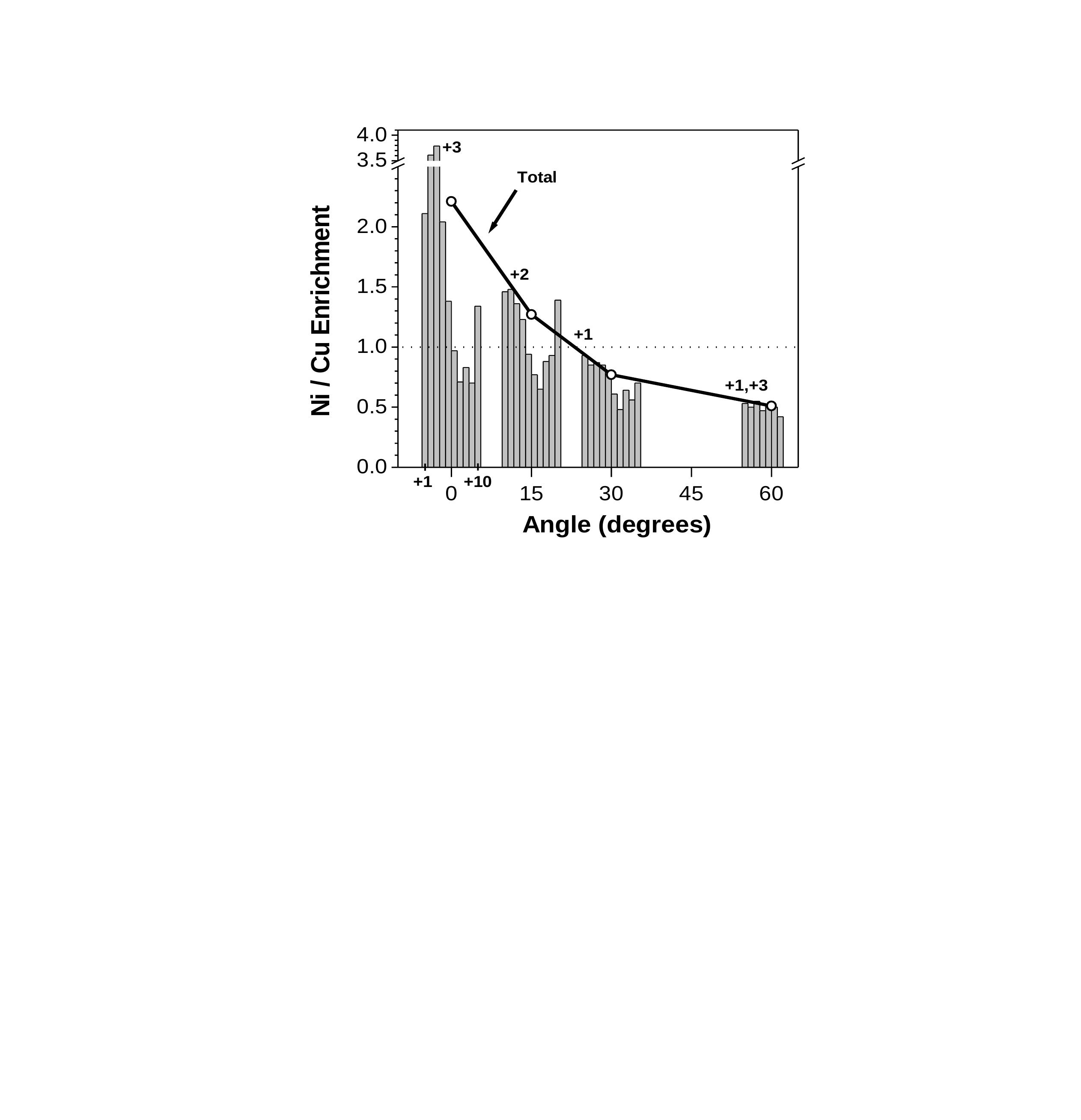}
   \vspace*{-11cm}
    \caption{Ni/Cu Enrichment Data as a Function of Angle}
    \label{fig:fig2_alt}
\end{figure}

\begin{figure}[H]
\vspace*{-13.30cm}
\hspace{-5.0cm}
  \includegraphics[width=1.5\textwidth, trim=0 0 0 0, clip]{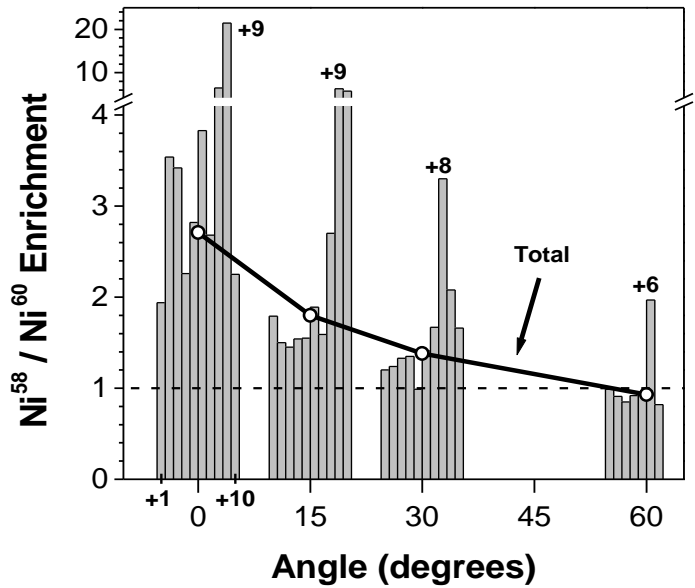}
  \vspace*{-5cm}
    \caption{Ni Isotope Enrichment Data as a Function of Angle}
    \label{fig:fig3}
\end{figure}

\begin{figure} [t]
  \vspace*{-17.0cm}
  \hspace{-5cm}
  \includegraphics[width=1.6\textwidth]{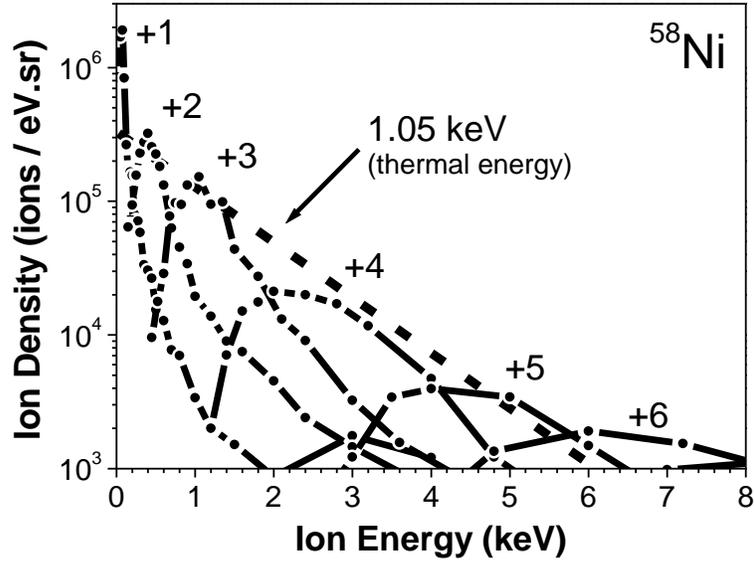}
  \vspace{-5cm}
  \caption{Ion Energy Distributions of  $ Ni^{58} $ as Function Charge State}
  \label{fig:fig4}
\end{figure}

Figure 4 presents ion density distributions, at zero angle, as a function of charge state and energy. These ion population data are used to extract a mean thermal energy to be used as the Maxwellian temperature in the ablative plasma.
Since the mean kinetic energy of an ideal gas is $(3/2)k_BT$, it is appropriate in the present case to replace $2k_BT$ in Eq. 2 with the measured mean kinetic energy of $1.05$ kV from figure 4. Further discussion of Figure 4 is presented in Appendix section 11.1.

\section{Experimental}

\subsection{Analysis of Data}
 Figure 5 shows a best fit of Equation 2 (i.e. Gaussian curve) to the 4 major data angles from Figures 2 and 3, with the additional requirement that the separation ratio be unity at the cross over point of 22.5 degrees for the Ni/Cu alloy case. This curve  is the result of combining the Ni/Cu data with the Ni isotope data, using average composite masses in Eq. 2, providing an overall averaged angular rotation rate of \(3.2 \times 10^9 \) rad/sec. 
The trend of the data points for both isotopes and alloy mimics the centrifuge curve across all angles, with the best fit dominating the enrichment region at angles below the cross-over point from enrichment to depletion in the alloy case. Analysis of these data points will now be done in terms of the specific centrifuge rotations that are expected to be involved.

\begin{figure}[htbp]
\raggedright
\vspace*{-1.0cm}
  \includegraphics[width= 0.9\textwidth, trim=0 0 0 0, clip]{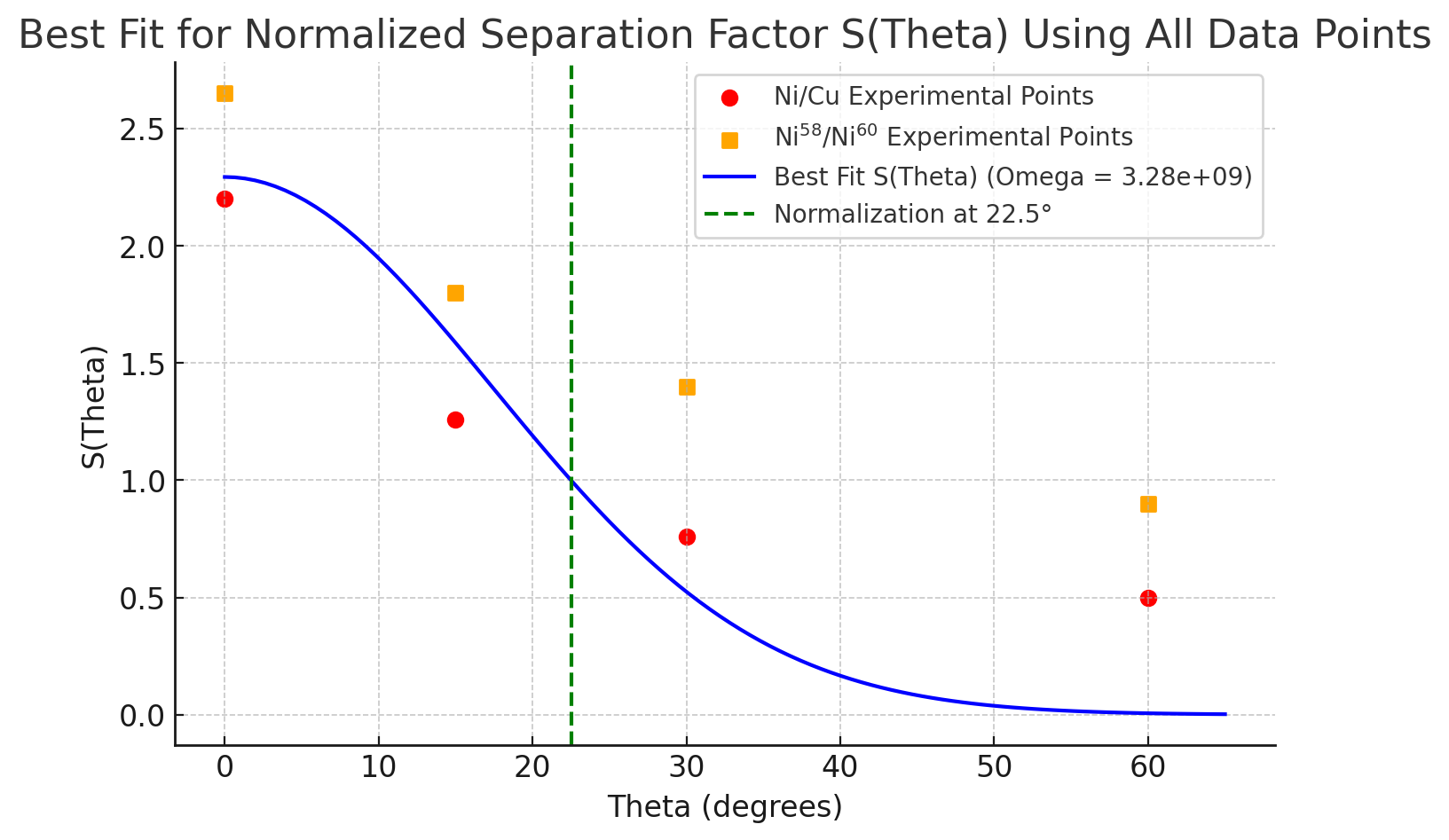}
    \caption{Best Fit for Normalized Separation Factor using Hybrid of Ion and Isotope Masses}
    \label{fig:fig6}
\end{figure}

\FloatBarrier

Rearranging Equation 2 the angular rotation rate, at each angle, can be expressed, using the experimental data points,in terms of the measured isotopic separation ratios, as:
\FloatBarrier
\begin{equation}
\omega(\theta) = \sqrt{\frac{2 k_B T \theta_{\max}^2 \left| \ln S(\theta) \right|}{\Delta m r_{\max}^2 \theta^2}}
\end{equation}
\\
which yields the specific angular rotation rates for each observed angle from Fig. 5
\footnote{It should be noted that $\omega(\theta)$ in Eq.~3 is model dependent since a choice of $r_{\text{max}}$ is required. We have estimated it at $125\,\mu\text{m}$ in all of our subsequent calculations. Choosing an $r_{\text{max}}$ of $150\,\mu\text{m}$ would  extend the centrifuge column but reduce the rotation rates by approximately $17\%$. Also, $\omega(\theta)$ in Eq. 3 is an effective rate containing both magnetic and electrostatic components to be discussed further in Sec. 7. Likewise the B values extracted in Sec. 3.2 below are effective values for the same reason. Algebraically these effective values fulfill the role of actual values in all subsequent calculations prior to Sec. 7}.

\begin{table}[h!]
\centering
\begin{tabular}{|c|c|c|}
\hline
\textbf{Angle \( \theta \)} & \textbf{\( \omega \) (rad/s), Ni/Cu} & \textbf{\( \omega \) (rad/s), Ni\textsuperscript{58}/Ni\textsuperscript{60}} \\
\hline
\( 0^\circ \)   & \( 4.50 \times 10^9 \)  & \( 12.7 \times 10^9 \) \\
\( 15^\circ \)  & \( 2.62 \times 10^9 \)  & \( 6.51 \times 10^9 \) \\
\( 30^\circ \)  & \( 1.43 \times 10^9 \)  & \( 2.46 \times 10^9 \) \\
\( 60^\circ \)  & \( 1.13 \times 10^9 \)  & \( 0.689 \times 10^9 \) \\
\hline
\end{tabular}
\caption{Effective rotation rates \( \omega(\theta) \) for Ni/Cu alloy and Ni\textsuperscript{58}/Ni\textsuperscript{60} isotope data at selected angles.}
\end{table}

Equation 3 has a singularity at zero angle; therefore, the values at theta equal to zero are obtained from a best-fit regression of the other 3 data points in each set that is extrapolated back to zero angle. These final equations and curves are presented below in Equations 4 and 5. Figure 6 shows the final fit of the quadratic and Gaussian equations to the rotation rate results and its extrapolation to the zero (center of the plasma column). Those curves are based on the following fitted equations which are used to extract magnetic fields as a function of observation angle $\theta$ (Section 3.2).
\begin{equation}
\omega_{\text{isotope}}(\theta) = 8.88 \times 10^9 \cdot \exp\left( -\frac{\theta^2}{26.79^2} \right) \quad \text{[rad/s]}
\end{equation}
\begin{equation}
\omega_{\text{alloy}}(\theta) = 1.63 \times 10^6 \cdot \theta^2 - 1.57 \times 10^8 \cdot \theta + 4.77 \times 10^9 \quad \text{[rad/s]}
\end{equation}
One of the things to notice about Figure 6 is that the isotope data points are systematically above the Ni/Cu data points as predicted by the cyclotron frequency equation. This is consistent with the cyclotron model as expressed by Eq. 3 in as much as the average alloy mass difference of 4.85 is significantly larger than that of the isotopes, which is 2. It predicts that the general or average rotation rate should scale as the inverse square root of the difference in mass for each case. Using a numerical average for each set of rates, one finds that the ratio of the average frequency for the alloy to that of the isotopes is 0.64 compared to the inverse square root of the mass differences which is 0.52 in general agreement with Equation 3.

\subsection{Plasma Rotational Dynamics}

In ultrafast laser ablation, the intense laser pulse, which itself has a Gaussian spatial profile \cite{diels2006ultrashort}, initially creates a highly ionized solid density plasma-plug at the point of impact with the target surface. Modeling this ablation zone as a cylinder with radius on the order-of or somewhat-larger-than the beam spot we can expect to have, within that zone: (a) strong electric fields due to charge separation and coulomb explosion, (b) strong, self-generated magnetic fields due to thermal and charge density gradients and hot electron transport, and (c) rapid plasma expansion transitioning from a dense cylindrical \textbf{}plasma to an expanding conical structure.

Within the initial cylinder there may exist two distinct types of rotational motion. One is the so called ExB drift rotation where the plasma undergoes bulk rotation as a whole. This results from the interaction between radial electric fields and axial magnetic fields. The entire plasma column undergoes rigid-body like rotation with angular rotation rate as given by \cite{chen2016introduction} \begin{equation}
    \omega_d = \frac{E_r}{R B_z}
\end{equation}
where $E_r$ is the radial electric field, $B_z$ is the axial magnetic field, and R is the rotating column radius. In this case, which is referred to as the rigid rotor model, the plasma is behaving as a hydrodynamic system rather than a particle plasma.
\\
\\

The other is cyclotron (Larmor) rotation involving individual charged particles moving around magnetic field lines. This results from the Lorentz force that compels electrons and ions into helical orbits. Each species rotates with their own cyclotron frequency \begin{equation}
    \omega_c = \frac{q B_{\text{eff}}}{m}
\end{equation}
which depends on mass m, charge q, and relevant magnetic field B leading to species separation, whereas the rigid rotor rotation relies on centrifugal potential energy acceleration to accomplish separation. 
Both modes of separation require a longitudinal magnetic field directed along the z axis of the rotating cylinder.

Equation 7 provides a way of extracting local magnetic fields, based on individual cyclotron rotation rates, as obtained from measured mass separation ratios. $B_{\text{eff}}$ represents an effective magnetic field extracted from the data as further extensively discussed in Section 7 below.

Shown below in Figure 6 are the results of using Eq. 7, for charge state 1 (which is the most abundant ion), to calculate the magnetic fields experienced by the alloy elements and the isotopes using their extracted rotation rates from Table 1 above. The numerical average taken from all the data points in Fig.6 provides an effective field value of 19.3 Megagauss [MG] within the ablation cylinder. This effective B field will be discussed further in Section 7 below where Ion Bernstein Wave contributions are presented.

\begin{figure}[t]
\raggedright
  \includegraphics[width= 1.1\textwidth, trim=0 0 0 0, clip]{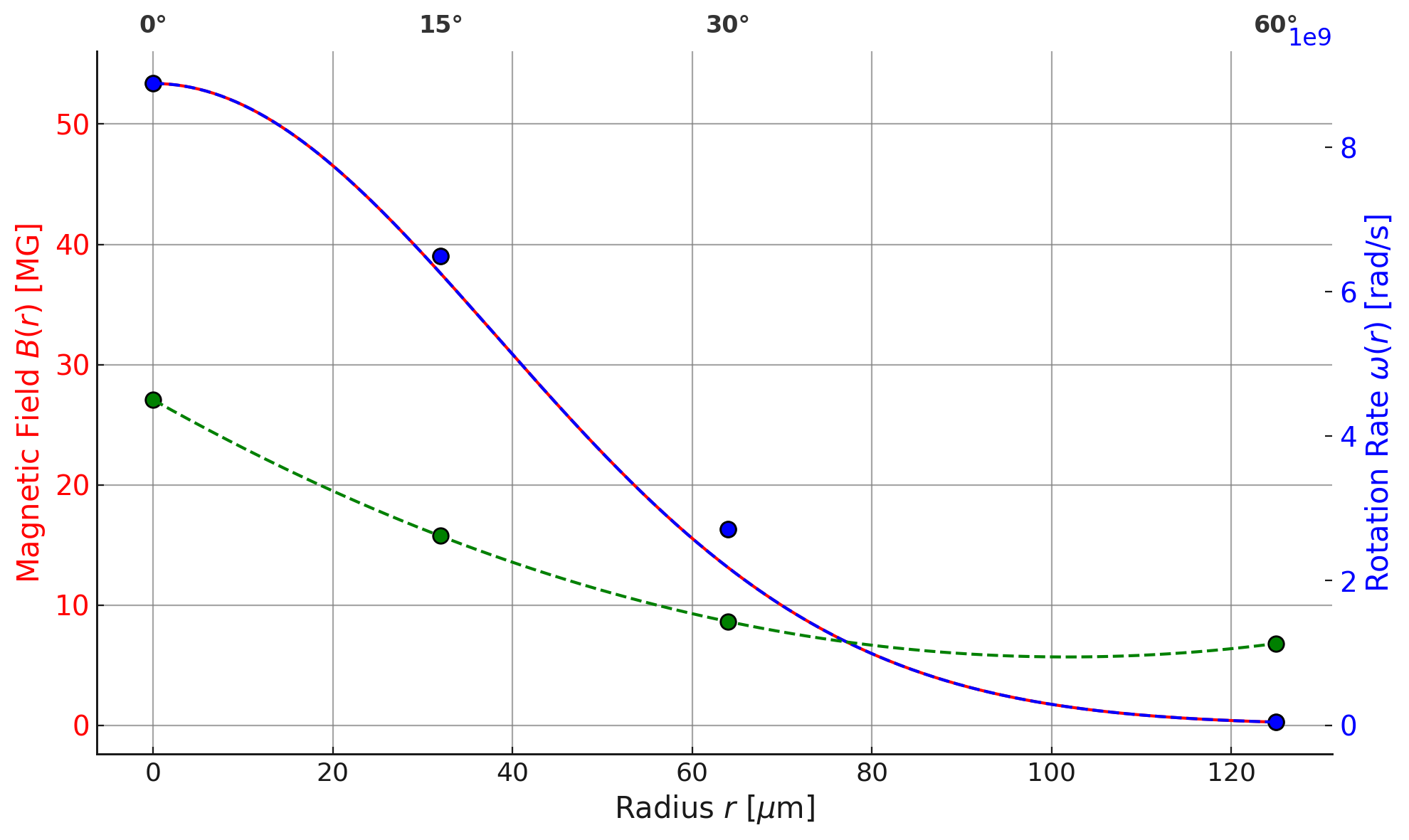}
    \caption{Gaussian (isotopes) and quadratic (alloy) fit to effective cyclotron rotation rates and magnetic field as function of plasma radius and observation angle.
    }

\end{figure}

The Gaussian equation that fits the local magnetic field for the isotope results as a function of the centrifuge cylinder radius is:

\begin{equation}
B(r) = 53.38 \exp\left( -\frac{r^2}{(55.82)^2} \right) \quad \text{[MG]}, \quad r \text{ in } \mu\text{m}
\end{equation}

and the quadratic equation used for the Ni/Cu alloy fit to the magnetic field data points is :
\begin{equation}
B(r) = 0.002259\, r^2 - 0.4538\, r + 28.64 \quad \text{[MG]}, \quad r \text{ in } \mu\text{m}
\end{equation}

Given this information about rotation rates, we can estimate the number of rotations an ion experiences in traveling the 125 $\mu$ length of our idealized centrifuge cylinder. We show, in Table 2 below, the distance traveled per rotation and the number of rotations achieved by an 0.2 keV Ni isotope ion traveling the length of the cylinder. It is seen in Figure 4 that a charge state one ion has this characteristic energy and represents the bulk of the ionic component in the plasma. The results in Table 2 are computed using separation ratios from Figure 3 and rotation rates from Eq.3. In Fig.3 the total enrichment ratios (for all charge states at a given angle) is shown as circles connected by a solid line. The distance traveled per rotation scales as $E^{1/2}$ for the higher energy ions with a commensurate decrease in number of predicted rotations. At 60 degrees (which has the slowest rotation rate), one full rotation would be accomplished by a 3.2 keV ion traveling the 125 microns. This represents a charge state between 4 and 5 according to Figure 9.

\begin{table}[h!]
\centering
\begin{tabular}{|c|c|c|c|}
\hline
$\theta$ (deg) & $\omega_c$ (rad/s) & Dist/Rot (m) & \# Rots in 125 $\mu$m \\
\hline
0   & $1.270 \times 10^{10}$ & $2.031 \times 10^{-6}$ & 61.54 \\
15  & $7.809 \times 10^{9}$  & $3.303 \times 10^{-6}$ & 37.84 \\
30  & $3.491 \times 10^{9}$  & $7.389 \times 10^{-6}$ & 16.92 \\
60  & $8.265 \times 10^{8}$  & $3.121 \times 10^{-5}$ & 4.01 \\
\hline
\end{tabular}
\caption{Cyclotron rotation rate, distance per rotation, and number of rotations to travel 125 µm for Ni ions at 0.2 keV.}
\label{tab:cyclotron_rotations}
\end{table}

\section{Analytical Model}
\subsection{Currents and Fields}
Let us concentrate on the isotope results in this section, leaving the alloy data for additional discussion at a later time.
Given that we now have a mathematical model for the magnetic field within the centrifuge cylinder for the isotopes, we can do the following:

1). Extract azimuthal current densities in the ablation plasma cylinder by using Ampere's law for the current density equation as a function of the measured magnetic fields.
2). Use this current density to extract the total circulating current that sustains the B(r) field and also to establish the electric field driving the magnetization current.
3). Use this driving electric field to extract a rotation rate for the rigid rotor centrifuge model.

\subsection*{Current Density}

Starting with the expression for the magnetic field $B_z(r)$ and using Ampère’s Law in cylindrical symmetry, and knowing that an azimuthal current density \( J_\theta(r) \) will generate an axial magnetic field \( B_z(r) \):
\begin{equation}
\oint \vec{B} \cdot d\vec{\ell} = \mu_0 I_{\text{enc}} \quad \Rightarrow \quad B_z(r) \cdot 2\pi r = \mu_0 \int_0^r J_\theta(r') \cdot 2\pi r' \, dr'
\end{equation}
(Note:- Theta in these equations is the azimuthal angle around the cylinder axis, not to be confused with the observation angle theta used to extract our data).
Differentiating both sides with respect to \( r \), results in the local form of Ampère’s Law:
\begin{equation}
\frac{dB_z(r)}{dr} = \mu_0 J_\theta(r)
\end{equation}

Using our experimentally determined form of Gaussian magnetic field profile:
\begin{equation}
B_z(r) = B_0 \exp\left(-\frac{r^2}{\sigma^2}\right)
\end{equation}

and differentiating:
\begin{equation}
\frac{dB_z(r)}{dr} = -\frac{2r B_0}{\sigma^2} \exp\left(-\frac{r^2}{\sigma^2}\right)
\end{equation}

which when substituted into Ampère’s law gives the azimuthal current density:
\begin{equation}
J_\theta(r) = -\frac{2r B_0}{\mu_0 \sigma^2} \exp\left(-\frac{r^2}{\sigma^2}\right)
\end{equation}

Equation 14 represents a differential current element in a ring segment of $2\pi r dr$ encircling the cylinder. Integrating this around the cylinder provides the total azimuthal current enclosed as a function of radius.

\begin{equation}
I_\theta(r) = \int_0^r J_\theta(r') \cdot 2\pi r' \, dr'
\end{equation}

Substituting for \( J_\theta(r') \), gives:
\begin{equation}
I_\theta(r) = -\frac{4\pi B_0}{\mu_0 \sigma^2} \int_0^r r'^2 \exp\left(-\frac{r'^2}{\sigma^2}\right) dr'
\end{equation}

The integral evaluates as:
\begin{equation}
\int_0^r r'^2 \exp\left(-\frac{r'^2}{\sigma^2}\right) dr' =
\frac{\sqrt{\pi}}{4} \sigma^3 \operatorname{erf}\left(\frac{r}{\sigma}\right) - \frac{r}{2} \sigma^2 \exp\left(-\frac{r^2}{\sigma^2}\right)
\end{equation} where erf is the error function.

\vspace{0.5 cm}
Substituting this into Eq. 16 result in the current as a function of radius:
\begin{equation}
I_\theta(r) = -\frac{\pi^{3/2} B_0 \sigma}{\mu_0} \operatorname{erf}\left(\frac{r}{\sigma}\right)
+ \frac{2\pi B_0 r}{\mu_0} \exp\left(-\frac{r^2}{\sigma^2}\right)
\end{equation}

    In the limit of large r, the enclosed current approaches the total effective current where $B_0$ is $B_{eff}$ at r = 0 and
\begin{equation}
I_{eff} = -\frac{\pi^{3/2} B_0 \sigma}{\mu_0}
\end{equation}

Figure 7 shows an overlay of the total current, current density, and cyclotron effective magnetic field as a function of the cylinder radius. It is seen there that the total effective current encompassing the cylinder is on the order of $1.3\times10^6$ amperes. This value will be further broken down in Section 7 where $B_{eff}$ is discussed.

\begin{figure}[t]
\raggedright
  \includegraphics[width= 0.8\textwidth, trim=0 0 0 0, clip]{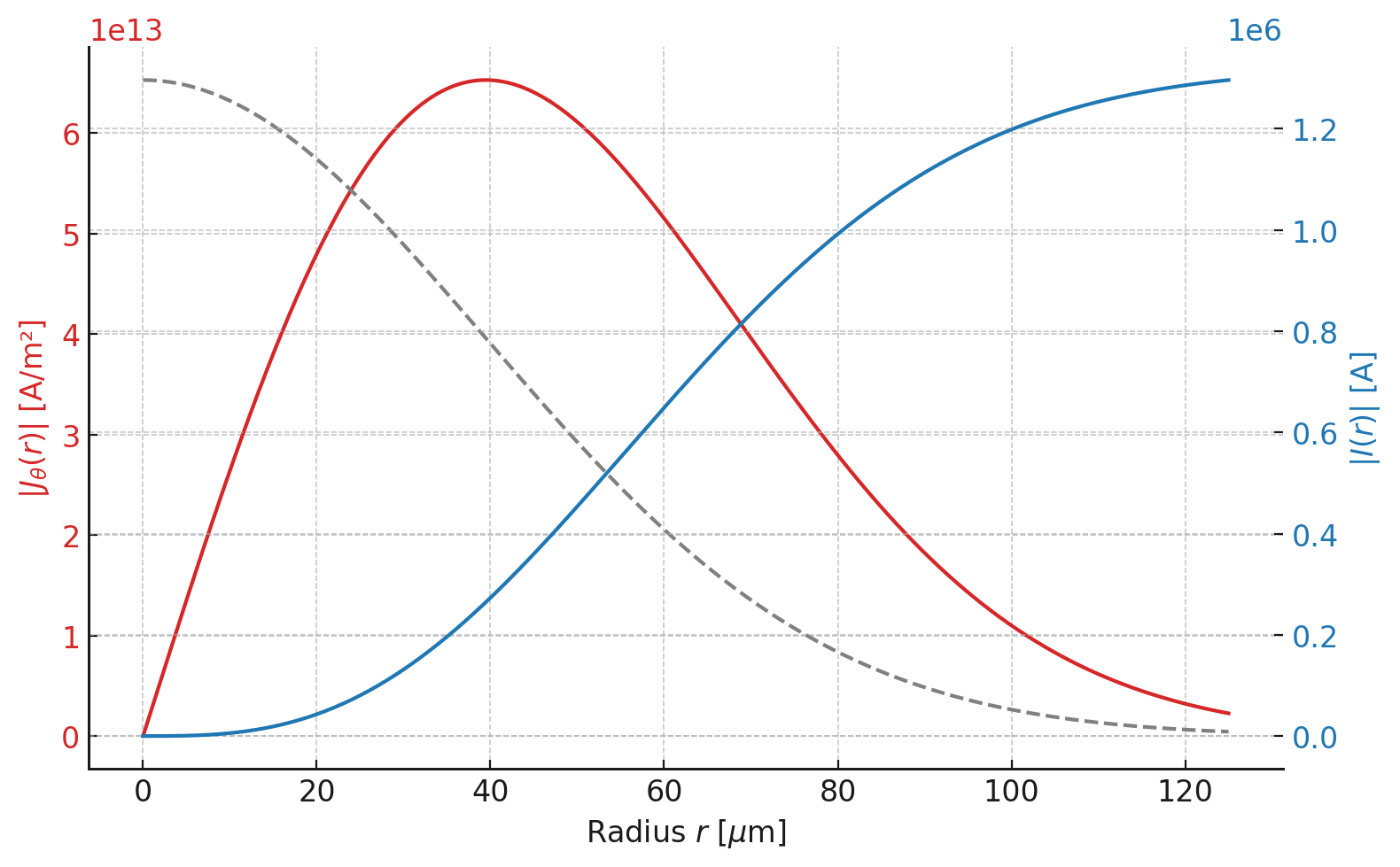}
    \caption{Total current (blue) surrounding the Gaussian magnetic field of the isotopes along with the extracted current density (red). Gray dashed line is the scaled magnetic field added for visual reference.}
\end{figure}

\section{Rigid Rotor Model}

In this section, we derive the rigid rotor rotation rate $\omega_r(r)$ for a cylindrically symmetric plasma column in which the azimuthal current density $J_\theta(r)$ generates a longitudinal magnetic field $B_z(r)$. The analysis assumes the plasma is well-conducting, axisymmetric, and governed locally by Ohm’s law in the azimuthal direction. In addition, by virtue of the current density, it is postulated that the electric field driving the azimuthal current also drives the $ E\times B$ drift.

We begin with Ohm’s law for the azimuthal electric field as discussed by Breginskii and Bettencourt for a magnetised plasma \cite{braginskii1965transport, bittencourt2013fundamentals}:
\begin{equation}
E_\theta(r) = \frac{J_\theta(r)}{\sigma_e}
\end{equation}
where $\sigma_e$ is the local plasma conductivity, and $J_\theta(r)$ is the azimuthal current density. As done previously, the current density is taken as the Gaussian-weighted linear form which was shown appropriate for the isotope field.
\begin{equation}
J_\theta(r) = -\frac{2 B_0}{\mu_0 \sigma^2} r \exp\left(-\frac{r^2}{\sigma^2}\right)
\end{equation}
where $B_0$ is the peak axial magnetic field at $r = 0$, $\mu_0$ is the vacuum permeability, and $\sigma$ is a characteristic radial scale length for the current and field profiles.

Substituting into Ohm’s law gives:
\begin{equation}
E_\theta(r) = -\frac{2 B_0}{\mu_0 \sigma_e \sigma^2} r \exp\left(-\frac{r^2}{\sigma^2}\right)
\end{equation}

The rigid rotor rotation rate $\omega_r(r)$ is defined as:
\begin{equation}
\omega_r(r) = \frac{E_\theta(r)}{r B_z(r)}
\end{equation}
Knowing that the axial magnetic field is also Gaussian:
\begin{equation}
B_z(r) = B_0 \exp\left(-\frac{r^2}{\sigma^2}\right)
\end{equation} and substituting $E_\theta(r)$ and $B_z(r)$ into the expression for $\omega_r(r)$  it is seen that $r$, $B_0$, and exponential terms cancel, leaving a constant:
\begin{equation}
\omega_r = -\frac{2}{\mu_0 \sigma_e \sigma^2}
\end{equation}

This result indicates that the plasma, for these field conditions, undergoes uniform rigid-body rotation at a fixed frequency, even though both $E_\theta(r)$ and $B_z(r)$ vary with radius. The rigidity arises due to the specific structure of the azimuthal current profile, which ensures the ratio $E_\theta(r) / (r B_z(r))$ remains constant.

The rotation rate depends on the plasma conductivity $\sigma_e$, determining the plasma’s response to electric fields, the magnetic permeability $\mu_0$, and the radial scale length $\sigma$, controlling the spatial distribution of the relevant fields. This model provides a self-consistent foundation for interpreting uniform rotation in ultrafast laser-produced plasmas with strong self-generated magnetic fields. Equation 27 provides the rotation rate needed to consistently sustain the current and field distributions in our self organized conductive plasma. In that regard, the fields are not independent inputs; they are outcomes of the plasma geometry and the conductivity and do not appear explicitly in the rigid rotor rotation equation. These fields are contained implicitly in the electron cyclotron rotation-rate and the electron ion collision frequency. 

Since the electrical conductivity is central to these results we provide, in Appendix 11.3, a brief discussion of the conductivity properties in an asymmetric magnetized plasma such as ours. It is shown there that the component of the conductivity tensor perpendicular to the magnetic field takes the form
\begin{equation}
\sigma_\perp = \frac{\sigma_0}{1 + \left( \frac{\omega_{ce}}{\nu_{ei}} \right)^2}
\end{equation}
where $\sigma_0$ is the conductivity in the absence of (or parallel to) the magnetic field, \( \omega_{ce} = \frac{eB}{m_e} \) is the electron cyclotron frequency, and \( \nu_{ei} \) is the electron-ion collision frequency. For sufficiently large values of $\frac{\omega_{ce}}{\nu_{ei}}$ the perpendicular component of the conductivity is diminished due to the action of the magnetic field continuously interrupting the forward velocity of the electrons between collisions with ions. It is shown there that, for our electron temperature and density, the electron ion collision frequency is very much larger than the the electron cyclotron frequency ($\nu_{ei} \sim 10^{18} \, \text{Hz}$, $\omega_{ce} \sim 10^{14} \, \text{rad/s}$). This means the electrons collide with ions well before they complete one full rotation resulting in $\frac{\omega_{ce}}{\nu_{ei}}$ essentially approaching zero. Thus, for our case, the perpendicular conductivity reduces to the field free Spritzer conductivity for a fully ionized plasma which is given by:
\begin{equation}
\sigma_e \approx 1.96 \times 10^7 \, \frac{T_e^{3/2}}{Z \ln \Lambda} \quad [\text{S/m}]
\end{equation}
where $T_e$ is in eV, $Z$ is the ion charge state, and $\ln \Lambda$ is the Coulomb logarithm. The computed value of this conductivity is found to be \[
\sigma_{\text{Spritzer}} = 1.96 \times 10^9 \, \text{S/m}
\] and when used with the gaussian scale length of 56 microns results in a rigid rotor rotation rate of\[
\omega_r = -2.59 \times 10^5 \, \text{rad/s}
\]
 The negative sign is because the rigid rotation is opposite to the sense of rotation for the cyclotron orbits. Parameters used for these calculations can be found in Appendix 11.3.

\subsection{Dominance of Cyclotron Rotation in Isotope Separation}

Our analysis of rotation rates of and within the plasma shows that ion cyclotron (Larmor) motion significantly dominates over collective $\vec{E} \times \vec{B}$ rigid rotor drift. Specifically, the measured rigid rotor frequency is four orders of magnitude lower than the corresponding ion cyclotron frequencies for the charge states of interest (see Table 1 above). As a result, while the $\vec{E} \times \vec{B}$ drift contributes to bulk plasma rotation at a global scale, it apparently plays a minimal role in isotope separation dynamics. Instead, separation is primarily governed by individual cyclotron-driven motion, which provides much stronger mass-dependent transport. This observation means that the key mechanism for isotope enrichment in our system arises from differences in Larmor rotation characteristics such as gyro radius and frequency rather than from hydrodynamic bulk processes. The implications are significant: mass-selective separation can occur efficiently provided that localized magnetic field forces are present to couple into the cyclotron trajectories of specific ion species both generally and resonantly.

\section{Resonant Isotope Enrichment- Ion Bernstein Waves}

The anomalously high isotope entrenchments for specific charge states, as seen in Figure 3 (and to a lesser extent in Figure 2), are suggestive of a resonance enhancement of the enrichment process. This effect can be observed for example in Fig.3 for the specific charge states of Ni-58 at charge states $ Z=9, 8, 6$ , and in particular being more dominant at smaller expansion angles. This behavior is consistent with and is, we propose, caused by resonant coupling between the cyclotron rotating ions and associated ion Bernstein waves (IBWs) that exist within a magnetized plasma \cite{swanson2020plasma, Ono_1993,stix1992waves}. IBWs are longitudinal electrostatic waves that propagate perpendicular to the magnetic field and exist in harmonics of the ion cyclotron frequency. These waves can interact resonantly with ions when their wave frequency matches the ion's cyclotron frequency or one of its harmonics. In our ultrafast laser ablation plasma, the strong, self-generated magnetic field produced in the early expansion phase enables a localized IBW spectrum that overlaps with the higher cyclotron frequencies of high charge state ions. Because these ions are also among the fastest particles in the plasma, they traverse the high-field region rapidly and are preferentially positioned to undergo coherent wave-particle coupling. This coupling leads to selective energy transfer and preferential transport of certain isotopes whose mass-to-charge ratio allows for near-exact resonance with the available IBW modes. The sharpness of these enhancements, and their confinement to specific charge states and angular observation regions, cannot be easily explained by other mechanisms such as thermal diffusion, bulk drift rotation, or other hydrodynamic  processes . Instead, they point to a nonlinear, field-mediated resonance mechanism such as IBW coupling, which can selectively amplify isotope populations far beyond what would be expected from normal equilibrium behavior. The presence of IBW-driven enrichment signatures, especially in low-density, high-$Z$ ion populations, thus serves as a compelling indicator of localized, resonance-based separation dynamics acting within the early-time magnetic structure of the plasma.

\subsection{Charge-State--Localized Resonance Condition:}

In this section we adopt a physical model in which a global IBW wave is created as a fundamental. Each charge state $Z$ then couples to this fundamental with a frequency equal to a harmonic of its local cyclotron frequency at its specific resonance radius $r_Z$. These local IBW frequencies are not all equal but are proportional to the charge state and the local magnetic field. With the cyclotron frequency being given as:
\begin{equation}
\omega_{c,Z}(r_z) = \frac{Z \cdot e \cdot B(r_z)}{m_i}
\end{equation}
the local expression for IBW matching is $\omega_{\text{IBW}}(Z) = \omega_{c,Z}(r_Z)$ and harmonics thereof. 

The dominant resonance in Figure 3 is the +9 charge state at the observation angle of zero degrees
where the isotope enrichment reaches a factor of 20 above natural abundance. We take this as the fundamental IBW frequency for the system. Other charge states with integer values of Z will have resonances at multiples of this fundamental frequency, and will occur at various radii, based on their location in the gaussian magnetic field. By calibrating to charge state $Z = 9$ at $r = 0$, a reference point is set up, whereby the fundamental IBW frequency is established for the system along with a set of harmonics. 
Using Eq.28 and matching the general cyclotron frequency to the fundamental IBW and its harmonics

\begin{equation}
\frac{nZ \cdot B(r_z)}{m_i} = \frac{9 \cdot B(0)}{m_i}
\quad \Rightarrow \quad
B(r_z) = \frac{9}{nZ} B(0)
\end{equation}
where $n \geq 1$ and $B(r_z)$ is given by our Gaussian field profile. Thus
\begin{equation}
    \frac{9}{nZ} B(0) = B(0) \cdot \exp\left( -\left( \frac{r_z}{\sigma_r} \right)^2 \right).
\end{equation}

Equation 29 defines the magnetic field needed to match the resonance conditions for each charge state and, by combining that with our field expression (Eq.8), a value of the radius is determined at which a particular Z charge state resonance should appear.

Canceling $B(0)$ on both sides of Eq. 30   and solving for $r$ gives the final expression for the resonance radius at which the charge states will appear:
\begin{equation}
r_z = \sigma_r \sqrt{ -\ln\left(\frac{9}{nZ}\right) }
 = \sigma_r \sqrt{ \ln\left(\frac{nZ}{9}\right)}
\end{equation}
Where $nZ \ge 9$. Also, since $n \geq 1$ the lower charge states will find resonances at r greater than zero and farther from the center, where the magnetic field is weaker. Eq. 31 has two variables, $\sigma_r$ and n. It is therefore necessary to assign harmonic numbers to the peaks observed from Fig. 3 in order to extract a $\sigma_r$ from Eq. 31 for comparison to that obtained from bulk isotopic enrichment ratios in Eq. 8 (and Fig. 6).

\subsection{Determination of Harmonic Order n and $\sigma_r$}
Using the global $Z=9$ anchoring, the fundamental IBW is
$\omega_0 = \Omega_{i,9}(0)$, and rearranging Eq. 30
the local harmonic number for charge state $Z$ at radius $r$ is
\begin{equation}
n(Z,r;\sigma_r)
=
\frac{9}{Z}
\exp\!\left(\frac{r^2}{\sigma_r^2}\right),
\label{eq:n_sigma}
\end{equation}
with the usual radial–angular mapping $r(\theta)=r_{\max}\frac{\theta}{60^\circ},
\quad
r_{\max}=125~\mu\mathrm{m}$.
To determine the Gaussian width $\sigma_r$ that best preserves low-order harmonic locking across the observed resonances, we minimize the total harmonic deviation
\begin{equation}
S(\sigma_r)
=
\sum_i
\left[
n(Z_i,r_i;\sigma_r)-N_i
\right]^2,
\label{eq:deviation}
\end{equation}
where $N_i$ are the target integer harmonics inferred from the data and $n(Z_i)$ the locally determined value. 
Consistent with the FLR-suppressed, low-order IBW picture, we assign (see Appendix Sec. 11.4)
\[
(Z,N) = 
(9,1)\ \text{at}\ 0^\circ,\quad
(9,1)\ \text{at}\ 15^\circ,\quad
(8,1)\ \text{at}\ 30^\circ,\quad
(6,2)\ \text{at}\ 60^\circ.
\]

Minimization of Eq.~(\ref{eq:deviation}) yields 
\begin{equation}
\
\sigma_r^{\star} \approx 2.4\times10^2~\mu\mathrm{m}
\end{equation}
as the Gaussian width that minimizes the total harmonic mismatch. Using the observed resonance angles from Fig. 3, and this optimized value of the Gaussian width, the computed $n_Z$ values are: 

\begin{table}[H]
    \centering
    \[ \boxed{ n_{9}(0^\circ)=1.000,\quad n_{9}(15^\circ)=1.017,\quad n_{8}(30^\circ)=1.204,\quad n_{6}(60^\circ)=1.968 } \]
    \caption{Calculated values of $n$ at various angles $\theta$}
    \label{tab:harmonic indices}
\end{table}

Thus showing that the highest charge state remains tightly locked to the fundamental harmonic inboard, while lower charge states approach low-order and slightly off resonance harmonics, progressively outward. This procedure provides a self-consistent method for extracting the effective magnetic-field width directly from the observed discrete resonance structure.

The integer target harmonics $N_i$ used in the minimization procedure were not chosen arbitrarily, but follow directly from the global $Z=9$ anchoring assumption and basic IBW coupling physics. See Appendix 11.4 for support of the above harmonic minimization approach. 
Higher-order harmonics (e.g., $n \gg 2$) are mathematically possible but are excluded by FLR suppression and by the experimentally observed low-order, discrete resonance structure.

These harmonic assignments are determined by (i) mathematical admissibility under the $Z=9$ anchor, (ii) lowest-harmonic preference from IBW coupling physics, and (iii) consistency with the observed radial ordering of resonances. So the harmonic numbers were not “fitted” —
they were constrained by the physics of harmonic accessibility and low-order IBW dominance. 
  We note that allowing arbitrarily high harmonic assignments ($n \gg 1$) would trivially reduce the least-squares deviation in Eq.~(\ref{eq:deviation}) by forcing it to match large integers at small radii through artificially small $\sigma_r$. Because $n$ grows exponentially with $r^2/\sigma_r^2$, decreasing $\sigma_r$ increases the harmonic number rapidly and permits spurious high-$n$ integer matches. Such solutions are physically inadmissible, as finite Larmor radius suppression strongly disfavors high-order harmonics in the ion Bernstein susceptibility. The minimization is therefore restricted to the lowest physically accessible harmonics consistent with FLR-dominated IBW coupling. As a further consistency test it will be observed that, for the charge state Z=6 case, where n is determined as 2, Eq. 31, using $\sigma_r$ as 240 microns, yields a predicted radial position for that charge state of 128.7 microns compared to its observed location of 125 microns.

\subsection{B Field Evolution 
Comparing Harmonic $\sigma^*_r$ and Enrichment $\sigma_r$}
The harmonic ordering of our resonance data resulted in a $\sigma^*_r$ value of 240 microns which is significantly greater than the Gaussian width obtained from the isotope enrichment analysis which was $\sigma_r$ = 56 microns. The introduction of these two different, but self similar, magnetic fields implies a certain time evolution of the relaxing magnetic field from its early initial state to the final quasi-equilibrium state. It is understood that the high charge state ions (e.g., $Z = 9, 8, 6$) produced in ultrafast laser ablation are typically those with the highest kinetic energies, originating from the peak of the Coulomb explosion (see Appendix 11.1). These ions travel at significantly higher velocities than the lower charge states and are therefore the first to traverse the magnetic field region immediately following plasma formation. This early-time field, generated through various initiating mechanisms and dynamo processes,  is characterized by a broader and more relatively uniform field distribution. It is during this initial phase that coherent resonant coupling with ion Bernstein waves (IBWs) is dominating with the high-$Z$ ions. These interactions leave distinct enrichment signatures that appear anomalous when compared to the behavior of the bulk plasma. As the plasma evolves, the initially broad field relaxes into our Gaussian distribution with $\sigma_r = 56\,\mu\mathrm{m}$ but retaining its peak value of $B_0 = 53.38\,\mathrm{MG}$ at the core of the cylinder. This temporal sequence supports a model in which early high-field interactions influence the motion of fast ions, while the later-stage, lower-field configuration governs the mass-selective separation dynamics of the slower,  more abundant, low-$Z$ species (See Figs. 4 and 9). In both cases the peak field at the core of the cylinder remains the same, with only the Gaussian width $\sigma_r$ changing.

Justification and support for collapsing the early field distribution to a narrower one can be found in the work of \cite{bell1993observation} where they demonstrate, with 2.1 picosecond near-infrared pulses, an azimuthal magnetic pinch effect collimating the plasma jet and focusing the ions into a half angle of 10 degrees about the normal to the ablating surface after laser pulse plasma formation. They demonstrate this both experimentally with time integrated x-ray images as well as computationally with a magneto hydrodynamic code demonstrating temperature and magnetic field contours about the normal direction with collimated jets of 100-to-240-micron length. Visual demonstration of such jet formation can also be found in \cite{albert2003time}.  In the present case therefore, the core of the plasma is retained intact, with the pinched peak field remaining the same, while the outer edges dissipate. Increasing $\sigma_r$ in this way does not require a higher peak field $B_0$, but instead simply reflects a magnetic field that more uniformly fills the centrifuge cylinder, prior to full pinching, providing a broader spatial variation of the magnetic field with radius. This initial field thereafter relaxes, through dissipative effects in the outer wings of the plasma, to evolve into the more narrow and concentrated quasi-equilibrium field that is responsible for the observed bulk isotope enrichment. See Fig. 8 below for a comparison of the shapes of these two fields. Further demonstration of such pinching can be found in \cite{borghesi1998megagauss}

The persistence and relaxation time of the two field types can be estimated by examining the ion transit times of the various fast and slow ions through the magnetic centrifuge cylinder. 
The fastest ions (e.g., charge state $Z=9$ to $6$) traverse the early, wide, high density Gaussian field.
These high-$Z$ ions are created at the peak of the Coulomb explosion and have the highest energy.
They move rapidly (tens of keV), reaching the end of the 125 micron plasma column in the $500$ to $800$ picoseconds range. These higher velocity ions are the ones being supported by ion Bernstein wave (IBW) resonances.
The slowest ions (e.g., $Z=1$, representing the bulk plasma) traverse the narrower, equilibrium Gaussian magnetic field.

These slower ions are more abundant and have lower energies ($\approx 0.2$ keV), resulting in longer transit times of a few nanoseconds.
They experience the relaxed magnetic field structure, which governs the bulk isotope separation.This sets the upper and lower bounds for the magnetic field creation and collapse time scales. Using the energy vs charge state data of Fig.9 (Sec. 11.1) for velocity information, the early magnetic field with wide gaussian must form on a very short time scale, significantly shorter than 537 picoseconds, which is the 125 micron transit time of the fastest +9 ion at 16.3 keV. This early field needs to last for a time of 835 picoseconds which is the transit time of the charge state +6 ion at 6 keV which is observed resonating at 60 degrees and is the slowest of the resonating ions. The final quasi-stationary field of the 56 micron Gaussian width needs to last for a time that is at least 4.85 nanoseconds to clear the transit time of the slowest low energy +1 ions at approximately 0.2 keV.

While high charge state ions (e.g., $Z = 9, 8, 6$) exhibit strong and selective resonant enhancements in our experimental data, it is important to recognize that these ions represent a very small fraction of the total plasma population being three to four orders of magnitude lower in density than the dominant low charge states ($Z = 1$ and $Z = 2$). Consequently, although the high-$Z$ species provide clear evidence of coherent wave-particle interactions, such as coupling to ion Bernstein waves (IBWs), they are unlikely to drive the bulk of the observed isotope separation. Instead, the primary separation process is governed by the dynamics of the much more abundant low charge state ions, which respond to the overall plasma structure. In this context, the high-$Z$ resonant enhancements serve primarily as diagnostic tracers of underlying plasma wave activity and magnetic field structure during the early stages of plasma evolution. They provide key information about the plasma's resonant behavior without being directly responsible for the majority of mass separation observed at later quasi-equilibrium stages. It is instructive to think about how these resonances could be brought down to the low Z ions in order to enhance energy coupling to them and the bulk of the mass enrichment process. Changing laser pulse parameters and plasma species suggests itself as possible ways to accomplish that. See Figures 6 and 8 in reference \cite{vanrompay2000isotope} as an example, where longer laser pulse durations are on the order of the cyclotron rotation period for Boron isotopes with associated enhancements of ionized states and enrichments.

\begin{figure}[H]
\raggedright
  \includegraphics[width= 0.75\textwidth, trim=0 0 0 0, clip]{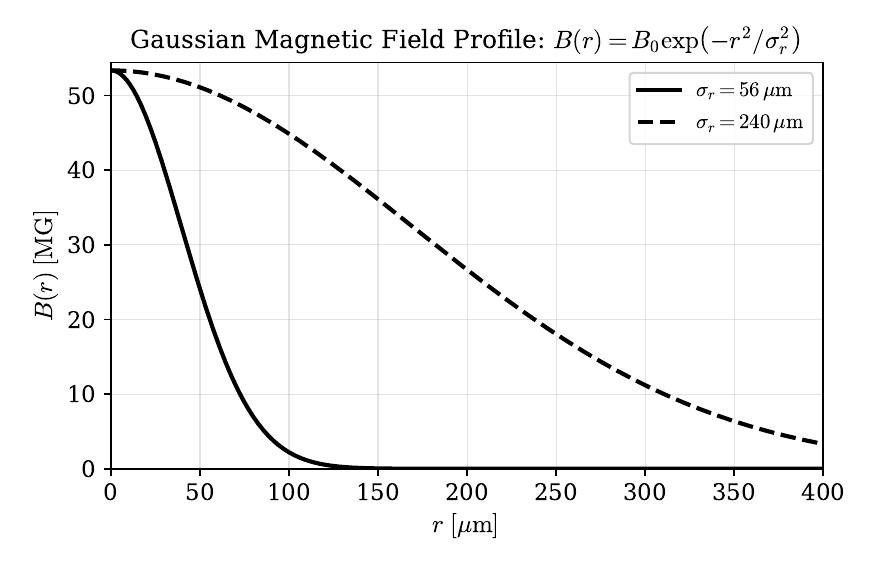}
\caption{Required effective magnetic field (dashed line) needed to satisfy the IBW enrichment resonances of high charge state (i.e. high velocity) ions compared to the field profile (solid line) used for the low velocity enrichment of the isotopes. Both curves are Gaussian but with differing width. (High velocity-240 micron, low velocity-56 micron).}
    \label{fig:fig6b}
\end{figure}

\section{Broad Spectrum IBW and $B_\mathrm{eff}$}

\
Having established the presence of IBW resonance waves in our ablation plasma, it is instructive to extend their role to the more generalized broad spectral case beyond the resonances of the previous section. Focusing on the late-time stable field condition and recognizing that since there is, in the plasma, a broad range of ion energies with different charge states (Fig. 4), along with a radially varying magnetic field (Fig.6), there will exist a dense and wide spectrum of cyclotron orbits with a similarly structured spectrum of IBW waves. As such, these spectral waves can drive near-resonant pumping across the entire distribution of ion cyclotron orbits. They will thereby produce larger values in the separation equation (Eq. 2) beyond that expected from the magnetic field rotation alone. Exact resonance isn’t required; proximity within the finite linewidth of each harmonic is enough. This gives a global, yet charge-selective, perpendicular heating and will greatly amplify isotope separation beyond the static B-field expectation alone.

As described in the previous section, an IBW wave is a longitudinal electrostatic wave whose k vector is perpendicular to the magnetic field and therefore aligned to transfer energy to the cyclotron orbits when sufficient frequency overlap occurs. These pulses of energy will increase the perpendicular orbital velocity by a delta increment upon each damping interaction. These velocity increments lead to an increased radial increment on the cyclotron orbits which results in an increase in isotopic fractionation. 

Looking at the centrifuge equation (Eq. 1), one sees that the enrichment ratio is exponentially dependent on both the rotation rate and the radius of rotation. This means there are two different channels by which enrichment can occur. The magnetic field controls the cyclotron ion rotation rate whereas the transverse velocity controls their radius of rotation, thereby resulting in two distinct enrichment mechanisms, one contingent on the existence of the other. If we assume only the B field is responsible for separations, when in fact an incremental radial contribution is also at play, then the extracted value of the B field will appear larger than its contribution is actually providing. This would appear to be the case for our originally extracted B field in Section 3.2. 

By way of example, consider the centrifuge plasma separation as being essentially a Boltzmann-in-a-rotating-frame, being expressed here as a sample case s, in a generalized cylinder of radius r as:
\[
n_s(r) = n_s(0)\,
\exp\!\left[
 \tfrac{\frac{1}{2} \Delta m_s \omega_r^2 r^2}{T_s}
\right].
\]
This relation shows a strong dependence on $r^2$ through the centrifugal term
$\tfrac{1}{2} \Delta m_s \omega_r^2 r^2$.
Now suppose an (IBW) resonance pushes just one species 
outward 
by a small radial increment $\Delta r$. Then the ratio of densities becomes
\begin{equation}
\frac{n_s(r+\Delta r)}{n_s(r)} 
\approx
\exp\!\left[
\frac{\Delta m_s\omega_r^2}{2 T_s}
\left(
2r\,\Delta r + \Delta r^2
\right)
\right].
\end{equation}

If the background rotation $\omega_r$ is large (as in our centrifuge case), 
even a modest $\Delta r$ yields a large exponential factor.
This explains the extraordinary separation values and large effective B field previously extracted. 
The IBWs only need to slightly nudge the broad spectrum of near-resonant species outward to achieve strong enrichment.

To this end, the extracted magnetic field B can be expressed as a linear combination of the actual longitudinal $B_z$ plus an additional electrostatic $B$\^*$_{ibw}$ to account for the total extracted B effective given as $B_{eff} = B_z + B$\^*$_{ibw}$.  

Although Ion Bernstein Waves (IBWs) are electrostatic modes and do not 
contribute directly to the quasi-static longitudinal magnetic field $B_{z}$, their presence profoundly alters the effective rotational dynamics inferred from the isotope-enrichment data. In our system, the centrifuge rotation is driven by the true axial field $B_{z}$, but the IBW spectrum---generated by mode conversion of electron plasma waves into ion Bernstein harmonics---produces 
additional azimuthal velocity and finite-radius excursions in the ions. These oscillatory IBW fields occur near integer multiples of the ion cyclotron frequency and therefore mimic the structure of a cyclotron spectrum, even though they do not supply additional axial magnetic flux. When the resulting ion distributions are interpreted using the standard centrifuge Boltzmann formalism, the combined cyclotron motion and IBW-driven excursions will appear as an enhanced effective exponential rotation term
$
\Omega_{\mathrm{eff}}^{2}
=
\Omega_{\mathrm{cf}}^{2}
+
\Omega_{\mathrm{IBW}}^{2}
$ where $\Omega_{\mathrm{cf}}^{2}$ is the ion cyclotron rotation rate and $\Omega_{\mathrm{IBW}}^{2}$ is the IBW contribution to the total effective rotation term. These two rotation contributions are added in quadrature since they appear as energy components in the Boltzmann exponent which requires the use of mean squared rotation factors. In this regard,
$\Omega_{\mathrm{eff}}$, can be understood 
as the result of an ``effective'' magnetic field
\begin{equation}
B_{\mathrm{eff}} \equiv \frac{m}{q}\,\Omega_{\mathrm{eff}}
\end{equation}
where $B_{\mathrm{eff}} = B_{z} + B$\^*$_{\mathrm{IBW}}$, not to imply the 
presence of a second magneto-static field, but as a convenient book-keeping 
device: $B$\^*$_{\mathrm{IBW}}$ denotes the equivalent magnetic-field strength that 
would produce, via ordinary cyclotron motion, the same additional mean-square 
azimuthal velocity and radial displacement that are actually produced 
electrostatically by the IBWs. This representation preserves the familiar 
centrifuge formalism while faithfully incorporating the wave--particle dynamics 
responsible for the observed enhancements in $\Omega_{\mathrm{eff}}$.

\subsection{$B_{\mathrm{eff}}$ Component Contributions}

It is constructive to now examine the way in which these two separation components contribute and interact. This is conveniently done by examining how different charge states contribute to the observed rotations within 
the plasma. For convenience, let us decompose the "effective" rotation frequency obtained from the original enrichment data using the system average taken over all detector observation angles. That hybrid rotation rate taken from Sec. 3.1 was  
\[
\Omega_{\mathrm{eff}} = 3.2\times10^{9}\ \mathrm{(rad/s)},
\]
Breaking it into two parts: (1) the true cyclotron rotation driven by the longitudinal 
magnetic field $B_z$, and (2) an additional contribution associated with the 
ion Bernstein wave (IBW) spectrum,

\begin{equation}
\Omega_{\mathrm{eff}}^{2}
=
\Omega_{\mathrm{cf}}^{2}
+
\Omega_{\mathrm{IBW}}^{2}
\end{equation}

the decomposition is defined by
\begin{equation}
\Omega_{\mathrm{cf}} = \frac{Z e B_z}{m},
\qquad
\Omega_{\mathrm{IBW}} = 
\sqrt{\Omega_{\mathrm{eff}}^{2} - \Omega_{\mathrm{cf}}^{2}},
\end{equation}
which yields a real IBW term whenever $\Omega_{\mathrm{cf}} \le \Omega_{\mathrm{eff}}$.  
This provides a practical criterion for determining which charge states can 
``participate'' in generating the observed effective rotation and at what longitudional field strength.

By way of reference, it is seen from Figure 3 that charge state 10 participates in the IBW resonance at 15 degrees. Thus, we can take that as the highest Z value present in the IBW affected plasma for our decomposition analysis. Using the equalities in Equation 38, the decomposition was performed for charge states $Z=1$ through $Z=10$ using various longitudinal magnetic field values for $B_z$. It was found that $B_z$ cannot exceed 1.9 MG in order for $\Omega_{\mathrm{IBW}}$ to remain real.

 \begin{table}[h!]
\centering
\begin{tabular}{c c c}
\hline
\textbf{Z} &
$\boldsymbol{\Omega_{\mathrm{cf}}}$ \textbf{(rad/s)} &
$\boldsymbol{\Omega_{\mathrm{IBW}}}$ \textbf{(rad/s)\^*} \\
\hline
1  & $3.16\times10^{8}$ & $3.18\times10^{9}$ \\
2  & $6.32\times10^{8}$ & $3.14\times10^{9}$ \\
3  & $9.48\times10^{8}$ & $3.06\times10^{9}$ \\
4  & $1.26\times10^{9}$ & $2.94\times10^{9}$ \\
5  & $1.58\times10^{9}$ & $2.78\times10^{9}$ \\
6  & $1.90\times10^{9}$ & $2.58\times10^{9}$ \\
7  & $2.21\times10^{9}$ & $2.31\times10^{9}$ \\
8  & $2.53\times10^{9}$ & $1.96\times10^{9}$ \\
9  & $2.84\times10^{9}$ & $1.47\times10^{9}$ \\
10 & $3.16\times10^{9}$ & $5.02\times10^{8}$ \\
\hline
\end{tabular}
\caption{Cyclotron and IBW "rotation" frequencies for maximum $B_{z}=1.9~\mathrm{MG}$ }
\end{table}

 The results, shown in  Table 4,
demonstrate several important trends. First, for this value of $B_z$, all charge 
states produce real IBW contributions, since $\Omega_{\mathrm{cf}}$ remains 
below $\Omega_{\mathrm{eff}}$ for the entire range $Z=1$ to $Z=10$. This means 
that every charge state examined contributes through some combination of cyclotron 
and IBW-driven motion.

Second, because $\Omega_{\mathrm{cf}} \propto Z$, the lower charge states have 
very small cyclotron frequencies and therefore require substantial IBW 
contributions to reach the observed rotation rate. For example, at $Z=1$ the 
cyclotron frequency is only $3.2\times10^{8}\ \mathrm{rad/s}$, nearly an order of 
magnitude smaller than $\Omega_{\mathrm{IBW}}$, so the IBW term supplies almost 
all the observed rotation for this species.

Third, as the charge state increases, the cyclotron term becomes a larger 
fraction of the total rotation. By $Z=10$ the cyclotron frequency has risen to 
$3.16\times10^{9}\ \mathrm{rad/s}$, approaching the observed effective frequency. 
The corresponding IBW frequency is smaller but still nonzero, indicating that 
even high charge states experience wave-driven motion.

Overall, the pattern follows directly from the relations in Equation 38.
As $Z$ increases, the cyclotron term grows linearly while the IBW term decreases 
accordingly, yet for $B_z = 1.9~\mathrm{MG}$ the IBW contribution remains real 
for all charge states considered. This result demonstrates that, under these 
conditions, all charge states from $Z=1$ to $Z=10$ can participate in generating 
the observed rotation through a combined action of true cyclotron motion and 
IBW-induced transverse dynamics. The lower charge states are dominated by the 
IBW component, while the higher charge states derive more of their rotation 
from the magnetic field but still retain a meaningful IBW contribution.

By way of completing the discussion on these combined enrichment channels, it should be mentioned that if, instead of using the generic system wide single frequency for demonstrating Eq. 38, we use the observed angularly dependent rotation rates from Table 1 for Ni isotopes, the overall outcome will be the same except that the results would produce a radially changing value of $B_z$ reflecting the Gaussian like nature of the data points, as shown in Table 5 for the Nickel isotopes. The numerical average here is 3.35 MG. 

\begin{table}[h!]
\centering
\begin{tabular}{c c}
\hline
\(\theta\) (deg) & \(B_{z,\max}\) (MG) \\
\hline
\(0^\circ\)  & 7.6  \\
\(15^\circ\) & 3.9  \\
\(30^\circ\) & 1.5  \\
\(60^\circ\) & 0.41 \\
\hline
\end{tabular}
\caption{Maximum longitudinal field consistent with a real IBW contribution at each observation angle for $\Omega_{eff}$ taken from Table 1 for Ni isotopes.}
\end{table}

There are relevant published reports that can be drawn upon for comparison to our experiment and extracted values of longitudinal magnetic field $B_z$. \cite{briand1985axial,sandhu2002laser, zhang2000hot}. The first involves normal incidence laser irradiation and the others use oblique incidence similar to the setup in this report. It can be argued that oblique incidence will produce a more pronounced longitudinal field compared to normal incidence due to tangential components of laser E field at the surface. Briand et al, using Faraday rotation and normal incidence, measure a longitudinal field of 0.6 MG in aluminum with a predicted theoretical maximum of 2.5 MG. Sandhu et al, using oblique incidence, measure ellipticity changes of a reflected probe pulse on aluminum and report a field with peak value of 27 MG droping to 5 MG over a 2 to 10 picoseconds time scale. Their B field is considered perpendicular to the probe propagation k vector. As such, due to oblique geometry, their determined B value will consist of both toroidal and longitudinal components. The relative contribution of each cannot be determined from their experiment, however their B field approaches 5 MG at long times of which 2.5 MG could be longitudinal. A third and perhaps the most relevant report is by Zhang et al where they examine the longitudinal magnetic field with Faraday rotation using laser parameters nearly identical to ours \cite{zhang2000hot}. They report a $B_z$ value of 1.76 MG. It can be concluded that our results for the longitudinal $B_z$ field are consistent with these references.

\section{IBW Effects on Radial Transition from Enrichment to Depletion}

At large observation angles (i.e. large centrifuge radii) the isotope data often do not show the expected transition from enrichment to depletion. For example in Figs. 2 and 3 it is seen that the alloy data shows clear transition from light ion enrichment to depletion between the 15 and 30 degree observation angles, whereas the isotope data does not. 

This behavior can be easily understood
from the IBW--centrifuge hybrid model and the relative percentage of IBW participation in each of the isotope separation examples. Because the IBW-driven radial displacement scales as (Eq. 35)
\[
\Delta r_{\mathrm{IBW}} \propto r(\theta),
\]
ions at large radius experience an enhanced effective centrifugal potential
even when the true cyclotron term becomes weak. The IBW contribution therefore
continues to act in the enrichment direction, preventing the usual large-radius
depletion that would occur in a pure magnetic centrifuge. The following is an explanation of how this happens.

% --- IBW modification of the centrifugal Boltzmann exponent (manuscript-ready) ---

\subsection{IBW modification of the centrifugal Boltzmann exponent}

In the pure magnetic-centrifuge limit, the isotope separation factor can be written in the Boltzmann form
\begin{equation}
S(\theta)\sim 
\exp\!\left[
-\frac{\Delta m}{m}\,
\frac{m\,\Omega_{cf}^{2}\, r_0(\theta)^{2}}{2T}
\right]
=
\exp\!\left[
-\frac{\Delta m}{2T}\,\Omega_{cf}^{2}\, r_0(\theta)^{2}
\right],
\label{eq:S_pure}
\end{equation}
where $r_0(\theta)$ is the guiding-center radius associated with the observation angle $\theta$.

With an IBW present, the instantaneous radius is written as
\begin{equation}
r(t)=r_0+\delta r(t),
\qquad
\delta r(t)\approx \Delta r_{\mathrm{IBW}}\cos(\omega t),
\end{equation}
where ions are subjected to electric field driven oscillation by the IBW waves with $\Delta r^2$ being the amplitude of these oscillations and $r_0$ the guiding center.

Recognizing that the centrifugal energy depends on the time-averaged mean-square radius of the cyclotron orbits, we examine
 the time average over many wave periods for 
\[
\langle r^2\rangle
=
r_0^2+2r_0\langle \delta r(t)\rangle+\langle \delta r^2(t)\rangle .
\] by evaluating the time average for each term selectively,
Since $\langle \cos(\omega_0 t)\rangle=0$ and 
$\langle \cos^2(\omega_0 t)\rangle=\tfrac12$, and
with $\langle \delta r\rangle=0$ for these symmetric oscillations the cross terms vanish  and

\begin{equation}
\langle r^{2}\rangle
=
\left\langle (r_0+\delta r)^2\right\rangle
=
r_0^{2}+\langle \delta r^{2}\rangle
=
r_0^{2}+\frac{1}{2}\Delta r_{\mathrm{IBW}}^{2},
\label{eq:r2avg}
\end{equation}

Using an IBW-driven displacement model (see Appendix Sec. 11.5)
\begin{equation}
\Delta r_{\mathrm{IBW}}
=
r_0\,\frac{\Omega_{\mathrm{IBW}}}{\Omega_{cf}},
\label{eq:drIBW}
\end{equation}
Eq.~\eqref{eq:r2avg} becomes
\begin{equation}
\langle r^{2}\rangle
=
r_0^{2}\left[
1+\frac{1}{2}\left(\frac{\Omega_{\mathrm{IBW}}}{\Omega_{cf}}\right)^{2}
\right].
\label{eq:r2avg2}
\end{equation}

Substituting $\langle r^{2}\rangle$ into the Boltzmann exponent gives the IBW--centrifuge hybrid form
\begin{equation}
S(\theta)\sim 
\exp\!\left[
-\frac{\Delta m}{2T}\,\Omega_{cf}^{2}\,\langle r^{2}\rangle
\right]
=
\exp\!\left[
-\frac{\Delta m}{2T}\,\Omega_{cf}^{2}\,r_0(\theta)^2
\left(
1+\frac{1}{2}\left(\frac{\Omega_{\mathrm{IBW}}}{\Omega_{cf}}\right)^{2}
\right)
\right].
\label{eq:S_hybrid}
\end{equation}

It is convenient to rewrite Eq.~\eqref{eq:S_hybrid} as a single ``effective rotation'' in the exponent by defining
\begin{equation}
\Omega_{\mathrm{eff}}^{2}
\equiv
\Omega_{cf}^{2}
\left[
1+\frac{1}{2}\left(\frac{\Omega_{\mathrm{IBW}}}{\Omega_{cf}}\right)^{2}
\right]
=
\Omega_{cf}^{2}+\frac{1}{2}\Omega_{\mathrm{IBW}}^{2},
\label{eq:Omeff_def}
\end{equation}
so that
\begin{equation}
S(\theta)\sim 
\exp\!\left[
-\frac{\Delta m}{2T}\,\Omega_{\mathrm{eff}}^{2}\,r_0(\theta)^{2}
\right].
\label{eq:S_omeff}
\end{equation}

Equations \eqref{eq:S_hybrid}--\eqref{eq:S_omeff} show explicitly that IBW-driven radial excursions enter the
centrifugal Boltzmann factor through $\langle r^{2}\rangle$, producing an additional $r_0(\theta)^2$-enrichment contribution from $\Omega_{IBW}$
that can persist at large observation angles and thereby preventing the expected transition from enrichment to depletion.

\vspace{1CM}
\section{IBW DRIVEN HARMONIC OSCILLATORS WITH DE-TUNING AND DAMPING}

Resonating IBW waves, as described in Sec. 6 above, modify the rotating cyclotron orbits through an electrostatic force describable, in its simplest form, as a driven harmonic oscillator with damping (de-tuning).
The classic differential equaiton for such a system is given as:
\begin{equation}
\ddot{x} + 2\gamma \dot{x} + \Omega_Z^{2} x
=
\frac{Z e}{m_i}\, E_{\perp}^{(\mathrm{IBW})}(r)\cos(\omega_0 t)
\label{eq:particle_oscillator}
\end{equation}

\noindent
where
$x(t)$ is the ion’s actual perpendicular displacement (e.g., the radial component of the orbit perturbation) about its unperturbed gyration trajectory; $\Omega_Z(r) = \dfrac{Z e B(r)}{m_i}$ is the local cyclotron frequency for charge state $Z$; $E_{\perp}^{(\mathrm{IBW})}(r)$ is the perpendicular ion Bernstein wave (IBW) electric field evaluated at the interaction radius;
$\omega_0$ is the IBW drive frequency (or dominant spectral component); and $\gamma$ the effective de-correlation or damping rate (collisions, dephasing due to spatial gradients, turbulent scattering, finite interaction time, etc.).

A consistent interpretation of Fig.~3 is that each observation angle corresponds to a dominant cyclotron--harmonic match for a particular charge state, but that the resonance itself is not infinitely sharp. Instead, it possesses a finite frequency width set by the intrinsic bandwidth of the ion Bernstein wave (IBW) packet and by nonlinear orbit distortion once the interaction becomes strong. This naturally explains why neighboring charge states are always present at a significant fraction of the dominant amplitude. For example, at $0^\circ$ the $+9$ state reaches roughly 20, while $+8$ still reaches about 6, or nearly $30\%$ of the dominant peak. 

If the resonance envelope from Eq. 47 is modeled with a simple Lorentzian response,
\[
A(\omega) \propto \frac{1}{\sqrt{(\omega - \Omega_Z)^2 + \gamma^2}},
\]
the observed amplitude ratio can be used to extract $\gamma$
\begin{equation}
\gamma
=
\frac{\Delta\Omega}
{\sqrt{\dfrac{1}{R^{2}} - 1}},
\qquad
R \equiv \frac{A(\Omega_Z+\Delta\Omega)}{A(\Omega_Z)}.
\end{equation} where $R<1$ since the amplitude at resonance is always larger than when off resonance. These relative amplitudes are read directly from Fig. 3

  $\Delta\Omega$ is the difference in rotational frequency of the two adjacent charge state  amplitudes being compared (e.g. 9 and 8 or 10 and 9, etc.). The IBW driven de-tuning factor is found to have, from Eq. 48, for the  Z ions involved, an effective Lorentzian half-width on the order of
\[
\frac{\gamma}{\Omega_Z} \sim 0.06,
\]
i.e.\ approximately $6\%$ of the cyclotron frequency. Such a width is large enough to allow adjacent charge states at a fixed B, whose cyclotron frequencies differ by approximately
\[
\frac{\Delta \Omega}{\Omega_Z} \simeq \frac{1}{Z} \approx 0.11,
\]
to remain well within the resonance band.

The observed charge-state mixing implies an effective resonance  full width at half maximum of
\[
\frac{2\gamma}{\Omega_Z} \sim 0.10\text{ -- }0.12,
\]
which is comparable to the intrinsic cyclotron-frequency separation between adjacent high-$Z$ charge states which is a fixed physical separation. ,
\[
\frac{\Delta \Omega}{\Omega_Z} = \frac{1}{Z} \approx 0.11
\quad \text{for } Z=9.
\]
This near-equality between the resonance full width and the charge-state splitting naturally accounts for the systematic overlap and mixing of neighboring charge states observed in Fig.~3.

In this view, the $15^\circ$ case, where $+9$ and $+10$ appear with comparable amplitude, is not anomalous but instead represents the situation in which the resonance envelope lies near the midpoint between two neighboring charge states. The broader pattern observed across all angles is therefore consistent with a finite-bandwidth IBW interaction centered on a dominant integer harmonic, with systematic charge-state mixing arising naturally from the resonance width rather than from local magnetic-field tuning.

\section {Conclusion}
An in-depth examination of ion dynamics associated with ultrafast laser ablation plumes, from $5.2 \times 10^{15}$  $W/cm^2$ pulses, emanating from nickel and a nickel/copper alloy surface provide detailed information about self-generated magnetic centrifuge fields in these ultrafast ablation plasmas. We examined, in particular, the effects of those fields on the radial mass enrichments around an axis normal to the ablating surface. It is shown that these energetic ablative ions, being subjected to such a self-generated and cylindrically symmetric longitudinal magnetic field oriented normal to the ablating surface, respond to a centrifuge mechanism that drives the radial distributions. Magnetic fields with  peak effective value of 53 MG and average effective fields of 20 MG are implied from isotope enrichment data with effective local azimuthal electric currents on the order of $1.3\times10^6$ amperes driving the fields.

These effective field values are further analyzed in terms of contributions from a longitudinal $B_z$ field plus an equivalent contribution from IBW electrostatic waves that greatly enhance the observed isotope enrichment and inflate these derived field values.

The analysis of our data, and our subsequent findings, proceeded as follows:
(1).	Obtain mass separation ratios by casting the standard centrifuge equation into the initial geometry of an ablating cylindrical model that, after a short distance, transitions to an expanding cone. Apply this model of an expanding cone to find angular distributions of observed mass enrichments in terms of separation ratios.
(2).These separation ratios are fitted to the centrifuge equation providing an overall averaged angular rotation rate for ions with an effective value of $3.2\times10^9$ radians/second. 
(3).The angular data scans are mapped back to the initial cylinder radial coordinate that allows detailed examination of the cyclotron orbits of ions at specific cylinder radii resulting in ion rotation rates at those radii. These are then converted to local magnetic field values, through the cyclotron equation, that provides a radial magnetic field distribution within the cylindrical centrifuge. This field distribution then provides a mathematical model that is used to extract current densities and enveloping currents responsible for the magnetic centrifuge mechanism.
(4).	The magnetic fields and currents are then used to construct the rigid rotor model of plasma rotation resulting in a value of $2.59\times10^5$ radians per second, significantly lower than the spatially averaged cyclotron rates. 
(5).	Anomalous localized charge state enrichments are examined in terms of resonant energy absorption from plasma waves that are identified as Ion Bernstein Waves (IBW). These are found to provide an good fit to the observed resonance data. A time dependent radial magnetic field profile is found necessary to properly describe these resonant data points. It is determined that an initial magnetic field lasts for at least 885 picoseconds and thereafter evolves to a stable quasi equilibrium centrifuge configuration that lasts for as long as 4.85 nanoseconds before dissipating.

The analysis is done using effective rotation rates for algebraic consistency and then subsequently resolving the results into longitudinal magnetic field $B_z$ and IBW components. Longitudinal magnetic fields of between 1.9 to 3.35 MG are extracted from the data  with between $4.6 \times 10^{4}$ to $8.2\times 10^{4}$ amperes for toroidal enveloping current sustaining this longitudinal field.

Resonant charge state isotope enrichment patterns are interpreted in terms of IBW driven harmonic oscillators with resonance de-tuning and damping. These patterns, observed across all angles, are consistent
with a finite-bandwidth IBW interaction centered on a dominant integer
harmonic, with systematic charge-state mixing arising naturally from finite resonance de-tuning relative to a global fundamental IBW wave.

Radial transitions from isotope enrichment to depletion are examined and explained in terms of Boltzmann weighting from Ion Bernstein Wave (IBW) contributions. The scope of this work shows that anomalously high Isotope enrichment in ultrafast laser ablation plasmas can be interpreted as driven by a centrifuge mechanism with individual ion cyclotron rotations within a  self generated magnetic field that is augmented by ion Bernstein wave (IBW) enhancements.

This work has relevance to applications in isotope harvesting and provides insight into laser plasma physics relative to direct ignition fusion research, especially the proton-Boron 11 programs and toroidal alpha channeling phenomena. In addition it demonstrates how isotopic analysis of laser plasmas can be used to derive a fuller understating of ablation phenomena, in particular its use in isotopically enriched thin film and nano-cluster deposition.

\section {Appendix}

\subsection{Coulomb Explosion}
As explained earlier, the ablation process under femtosecond laser ablation occurs by a coulomb explosion due to the lighter electrons being stripped from the surface leaving a high concentration of positive ions in a solid density plasma condition. We described this explosion, in a previous report in terms of a parallel capacitor electric field model \cite{vanrompay1998pulse}. That model predicts a linear relationship between the charge state and its energy. The data below show that a quadratic behavior is more typical and needs to be analyzed in those terms. See Figure 4 for details where the data points in Figure 9  are taken as the most probable energy at each charge state.

\begin{figure}[H]
\raggedright
  \includegraphics[width= 0.75\textwidth, trim=0 0 0 0, clip]{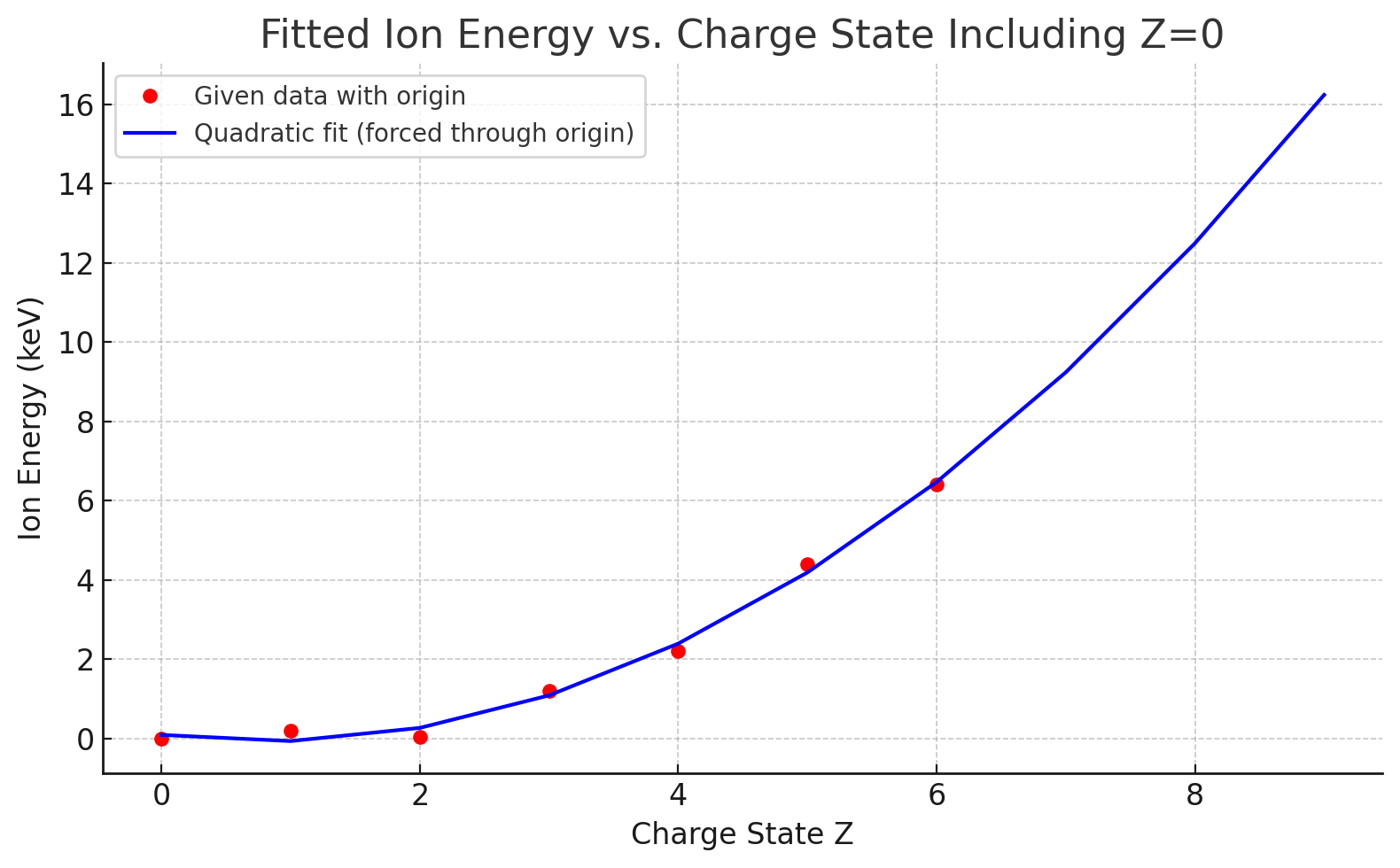}
    \caption{Mean ion energy as function of charge state shows super liner (quadratic) coulomb explosion. See Fig.4 for details.}
    \label{fig:fig6b2}
\end{figure}
These points can be fit nicely to a quadratic equation passing near the origin.
\begin{equation}
E(Z) = 0.244\,Z^2 - 0.399\,Z + 0.094 \quad \text{[keV]}
\end{equation}

This result can be interpreted as higher charge states experiencing stronger net forces during the coulomb explosion suggesting that higher $Z$ ions originate from deeper within the initially formed solid density plasma, where the local positive potential is strongest, thus experiencing the largest forces and gain the most energy. The non-linearity reflects a collective effect not just from its own charge, but through interaction with the entire charge cloud.

In a Coulomb explosion, each ion is repelled by the net positive charge of its neighboring ions. The resulting force on an individual ion scales approximately as:

\[
F \propto Z \cdot E_{\text{local}}
\]

where $Z$ is the ion charge state and $E_{\text{local}}$ is the local electric field generated by the surrounding ion cloud. Importantly, this field is not constant—it reflects the cumulative contribution of neighboring ions, which also depend on their respective charge states $Z_j$.

The energy gain of an ion depends on both its own charge and the electric potential it experiences. This can be approximated as:

\[
E_i \propto Z \cdot \Phi_{\text{Coulomb}} \propto Z \cdot \left( \sum_j \frac{Z_j}{r_{ij}} \right)
\]

where $r_{ij}$ is the distance between ion $i$ and neighboring ion $j$.

In a dense central region of the exploding plasma, the potential $\Phi_{\text{Coulomb}}$ scales linearly with the total enclosed charge, leading to a quadratic type dependence of energy on charge state:

\[
E_i \propto Z^2
\]

This relationship aligns with experimental observations showing that higher-$Z$ ions emerge with disproportionately larger kinetic energies. It reflects the collective nature of the explosion, where the energy is not solely a function of individual charge, but of coherent interaction with the entire charge distribution.

\subsection{Collimation Restrictions}

The sector analyzer has a 2 mm acceptance aperture at its input \cite{van2003mass, vanrompay2000isotope}. The angular resolution \( \Delta \theta \) of this ion detector is the half-angle of the acceptance cone subtended by the detector aperture as seen from the target:

\[
\Delta \theta = \tan^{-1} \left( \frac{r_{\text{aperture}}}{d} \right)
\]

Where:
\( r_{\text{aperture}} \)= radius of the detector aperture and
\( d \) is the distance from the target to the detector
\[
r_{\text{aperture}} = \SI{1.0e-3}{\meter}
\quad\text{and}\quad
d = \SI{1.1}{\meter}
\]
\[
\Delta \theta = 
\approx 9.09 \times 10^{-4} \, \text{rad = } \ang{0.0521}
\]

Using the angle to radius conversion from Fig.1

\[
\frac{r}{r_{\text{max}}} = \frac{\theta}{\theta_{\text{max}}}
\quad \Rightarrow \quad
r = r_{\text{max}} \cdot \frac{\theta}{\theta_{\text{max}}}
\]

\[
r = 1.25 \times 10^{-4} \cdot \frac{0.052}{60}
= 1.083 \times 10^{-7}m \
\]

Only ions emitted from within a differential radial zone of \( \sim 0.108 \, \mu\text{m} \) will fall within the detector's acceptance cone at any particular observation angle. The detector has a circular view of the ablation plume with diameter of\( \sim 0.217 \, \mu\text{m} \) at any fixed observational angle and therefore acts as an ion ablation microscope.

\subsection{Conductivity Perpendicular to B}
\section*{Appendix: Perpendicular Conductivity in a Magnetized Plasma}

Begin with the steady-state force balance equation for electrons in a magnetized, collisional plasma \cite{boyd2003physics, bittencourt2013fundamentals}:
\[
e \vec{E} =  - e \vec{v}_e \times \vec{B} - m_e \nu_{ei} \vec{v}_e
\]
This equation reflects the balance between the electric force, the Lorentz force from the magnetic field, and the collisional drag acting on electrons due to scattering from ions. Assuming linear response, this equation determines the steady-state electron drift velocity \( \vec{v}_e \) in terms of the applied electric field \( \vec{E} \).

To solve this equation, we choose a coordinate system where the magnetic field is aligned along the \( \hat{z} \)-axis, i.e., \( \vec{B} = B \hat{z} \), and write the drift velocity as \( \vec{v}_e = (v_x, v_y, v_z) \). The cross product \( \vec{v}_e \times \vec{B} \) then becomes \( (v_y B, -v_x B, 0) \), and the collisional drag term is simply \( \nu_{ei} \vec{v}_e \), where \( \nu_{ei} \) is the electron-ion collision frequency.

Substituting into the force balance and equating components yields:
\[
\begin{aligned}
-E_x &= \nu_{ei} v_x + \omega_{ce} v_y \\
-E_y &= \nu_{ei} v_y - \omega_{ce} v_x \\
-E_z &= \nu_{ei} v_z
\end{aligned}
\]
where \( \omega_{ce} = eB/m_e \) is the electron cyclotron frequency. Solving this system for \( \vec{v}_e \) in terms of \( \vec{E} \) leads to the expressions:
\[
\begin{aligned}
v_x &= \frac{-\nu_{ei} E_x - \omega_{ce} E_y}{\nu_{ei}^2 + \omega_{ce}^2} \\
v_y &= \frac{-\nu_{ei} E_y + \omega_{ce} E_x}{\nu_{ei}^2 + \omega_{ce}^2} \\
v_z &= -\frac{E_z}{\nu_{ei}}
\end{aligned}
\]

From these, the current density \( \vec{J} = -e n_e \vec{v}_e \) is computed, yielding a linear relation between \( \vec{J} \) and \( \vec{E} \) through a conductivity tensor:
\[
\vec{J} = \boldsymbol{\sigma} \cdot \vec{E}
\]

The resulting conductivity tensor takes the form:
\[
\boldsymbol{\sigma} =
\sigma_0
\begin{bmatrix}
\frac{1}{1 + b^2} & -\frac{b}{1 + b^2} & 0 \\
\frac{b}{1 + b^2} & \frac{1}{1 + b^2} & 0 \\
0 & 0 & 1
\end{bmatrix}
\quad \text{with} \quad b = \frac{\omega_{ce}}{\nu_{ei}}
\]

Here, $\sigma_0$  is the classical Spitzer conductivity (for \( \vec{B} = 0 \)), and \( b \) is the dimensionless magnetization parameter.

The diagonal entries of the tensor describe the conductivity \emph{parallel} and \emph{perpendicular} to the magnetic field. The off-diagonal elements represent the \emph{Hall conductivity}, which accounts for the rotation of the current vector due to the Lorentz force. Specifically, the conductivity perpendicular to the magnetic field is given by:
\[
\sigma_\perp = \frac{\sigma_0}{1 + \left( \frac{\omega_{ce}}{\nu_{ei}} \right)^2}
\]

This result shows that the presence of a magnetic field suppresses the ability of electrons to carry current across field lines. In the limit of strong magnetization (\( \omega_{ce} \gg \nu_{ei} \)), the perpendicular conductivity becomes significantly smaller than the un-magnetized case. In contrast, the conductivity parallel to the field remains unaffected and equals \( \sigma_0 \).
It is seen that the components of the tensor are controlled by the electron cyclotron frequency and the electron-ion collision rate. These are, in turn, controlled by the electron density, electron plasma temperature, and magnetic field strength.

\begin{table}[h!]
\centering
\caption{Key Assumptions for Conductivity Calculation in 125 µm Plasma Cylinder}
\footnotesize
\begin{tabular}{|l|c|}
\hline
\textbf{Parameter} & \textbf{Assumed Value / Condition} \\
\hline
Plasma geometry & Cylindrical, radius $\sim 125 \, \mu$m \\
Plasma species & Fully ionized Ni or Ni/Cu alloy \\
Charge state & $Z = 1$ \\
Electron temperature & $T_e = 100 \, \text{eV}$ \\
Electron density & $n_e \sim 10^{24} - 10^{26} \, \text{m}^{-3}$ \\
Coulomb logarithm & $\ln \Lambda = 10$ \\
Magnetic field strength & $B(\theta) \leq 5338 \, \text{T}$ (Gaussian profile) \\
Plasma regime & Fully ionized, weakly coupled, collisional \\
Conductivity model & Spitzer conductivity with perpendicular suppression \\
Conductivity formula & 
$\displaystyle \sigma_\perp = \frac{1.96 \times 10^7 \, T_e^{3/2}}{Z \ln \Lambda \left[1 + \left(\frac{\omega_{ce}}{\nu_{ei}}\right)^2\right]}$ \\
\hline
\end{tabular}
\label{tab:conductivity_assumptions}
\end{table}

Using the empirical formula for the electron-ion collision frequency \cite{boyd2003physics} (with $T_e$ in eV, $n_e$ in m$^{-3}$):

\[
\nu_{ei} \approx 2.91 \times 10^{-6} \cdot \frac{n_e Z \ln \Lambda}{T_e^{3/2}} \quad \text{[Hz]}
\]
it is found that 
\[
\nu_{ei} \sim 2.8\times10^{18} \, \text{Hz}
\]
and more specifically for our isotope gaussian field distribution:

\[{1.10 \times 10^{14} \, \text{rad/s} \leq \omega_{ce} \leq 9.39 \times 10^{14} \, \text{rad/s}}
\]
Based on these electron-ion collision frequencies it is seen that the perpendicular conductivity reduces to the field free Spitzer value.
\[
\sigma_0 \approx \frac{1.96 \times 10^7 \, T_e^{3/2}}{Z \ln \Lambda} \quad \text{[S/m]}
\]

\subsection{Selection of Target Integer Harmonics}

The harmonic assignments used in the minimization procedure are not arbitrary; they follow directly from the structure of the Ion Bernstein Wave (IBW) resonance condition combined with the assumed Gaussian magnetic field profile.

We begin with the resonance relation
\begin{equation}
\omega_0 = n\,\Omega_{i,Z}(r),
\qquad
\Omega_{i,Z}(r) = \frac{Z e B(r)}{m_i},
\end{equation}
and anchor the global frequency at the dominant central resonance, $\omega_0 = \Omega_{i,9}(0)$.

This yields the predicted harmonic number for each charge state (from Eq. 30) as:
\begin{equation}
n(Z,r;\sigma_r)
=
\frac{\Omega_{i,9}(0)}{\Omega_{i,Z}(r)}
=
\frac{9}{Z}
\exp\!\left(\frac{r^2}{\sigma_r^2}\right),
\end{equation}
where our usual Gaussian magnetic field has been employed.

Two structural consequences follow immediately:

\begin{enumerate}
\item \textbf{Charge scaling:} In the limit $\sigma_r \to \infty$ (spatially uniform field), the harmonic number approaches $n \to \frac{9}{Z}$.

Thus lower charge states naturally correspond to higher effective harmonic numbers even without radial field decay.

\item \textbf{Radial amplification:} Because the Gaussian magnetic field decreases with radius, the exponential factor increases $n$ with increasing $r$, producing a natural outward harmonic ladder.
\end{enumerate}

\paragraph{Z = 9 at $0^\circ$ and $15^\circ$.}
At $r=0$, $n(9,0) = 1$,
independent of $\sigma_r$. 

The central resonance is therefore uniquely and necessarily assigned $n=1$. 
At $15^\circ$, the radius remains small and the exponential factor remains close to unity; thus the physically consistent assignment remains $n=1$.

\paragraph{Z = 8 at $30^\circ$.}

For $Z=8$,
$n \to \frac{9}{8} = 1.125
\quad \text{as} \quad \sigma_r \to \infty$.

Even in the absence of radial decay the harmonic number already exceeds unity. At finite radius it increases further. The nearest low integer harmonic is therefore $n=1$, with the understanding that this case may be modestly off resonance. Assigning $n=2$ would require an unrealistically small $\sigma_r$ and would disrupt the coherence of the global harmonic structure.

\paragraph{Z = 6 at $60^\circ$.}
For $Z=6$,
$n \to \frac{9}{6} = 1.5
\quad \text{as} \quad \sigma_r \to \infty$.

Thus $n=1$ is mathematically impossible at this location, since the baseline value already exceeds unity before any radial amplification is included. Because the exponential factor increases $n$ with radius, the nearest accessible integer above the baseline value of $1.5$ is
$n = 2$. Higher integers (e.g.\ $n=3$) would require an excessively small magnetic scale length and would force the remaining charge states far from near-integer behavior. Therefore $n=2$ is the lowest integer that is both mathematically allowed and globally consistent.

\paragraph{Final Harmonic Set.}

The integer targets used in the least-squares minimization are therefore
\begin{align}
Z=9,\; 0^\circ &\rightarrow n=1, \\
Z=9,\; 15^\circ &\rightarrow n=1, \\
Z=8,\; 30^\circ &\rightarrow n=1, \\
Z=6,\; 60^\circ &\rightarrow n=2.
\end{align}

This assignment reflects: (i) anchoring of the dominant central resonance, (ii) monotonic outward harmonic increase for decreasing $Z$, (iii) mathematical constraints of the resonance formula, and (iv) preservation of a coherent global IBW harmonic ladder.

\subsection{Radial Extension of Enrichment Values}

In the hybrid model, the magnetic confinement defines a characteristic
azimuthal velocity scale
\begin{equation}
v_\theta = \Omega_{\mathrm{cf}} r_0,
\qquad
\Omega_{\mathrm{cf}} = \frac{q B_z}{m},
\label{eq:vtheta_eff}
\end{equation}
where $\Omega_{\mathrm{cf}}$ is the true cyclotron frequency associated
with the longitudinal magnetic field.

Ion Bernstein waves introduce an additional transverse velocity
perturbation $\delta v_\perp$ through their oscillatory electrostatic
field. Rather than interpreting this perturbation as a physical rotation,
we parameterize it in angular form by defining an effective
IBW velocity scale,
\begin{equation}
\delta v_\perp \equiv \Omega_{\mathrm{IBW}}^{(\mathrm{eff})}\, r_0,
\label{eq:omega_ibw_eff_def}
\end{equation}
where $\Omega_{\mathrm{IBW}}^{(\mathrm{eff})}$ is an energy-equivalent
parameter with units of rad/s, but does not represent a true rotation.

A transverse velocity perturbation modifies the canonical angular
momentum and therefore the guiding-center radius. To leading order,
the fractional radial excursion scales as
\begin{equation}
\frac{\Delta r_{\mathrm{IBW}}}{r_0}
\sim
\frac{\delta v_\perp}{v_\theta}.
\label{eq:dr_ratio}
\end{equation}

Substituting Eqs.~(\ref{eq:vtheta_eff}) and~(\ref{eq:omega_ibw_eff_def})
gives
\begin{equation}
\frac{\Delta r_{\mathrm{IBW}}}{r_0}
\sim
\frac{\Omega_{\mathrm{IBW}}^{(\mathrm{eff})}}{\Omega_{\mathrm{cf}}}.
\label{eq:dr_final}
\end{equation}

Equation~(\ref{eq:dr_final}) expresses IBW-driven radial broadening
as the ratio of the IBW-induced transverse velocity scale to the
magnetic confinement velocity scale. As $\Omega_{\mathrm{cf}}$
decreases with radius due to the radial profile of $B_z$, the same
IBW-induced perturbation produces a larger fractional radial excursion,
causing IBW effects to become increasingly important at large
observation angle.

\section*{Acknowledgments}
The experimental work reported here was performed at the Gerard Mourou Center for Ultrafast Optical Science, University of Michigan, Ann Arbor, MI during the years 1999-2002.
The authors would like to thank the University of Michigan and Gerard Mourou, while he was director at the Center, for providing support and resources during the experimental phase of this work. Technical assistance and laboratory support was provided by Zhiyu Zhang and John Nees.This work was supported in part by the National Science Foundation through the Center for Ultrafast Optical Science under Grant No. STC PHY 8920108, and by the Airforce Office of Scientific Research through an AFOSR DURIP equipment Grant No. F49620-95-1-0474  
\bibliographystyle{unsrtnat}
\bibliography{references}

\end{document}